\documentclass[showpacs,prb,aps,superscriptaddress,twocolumn,floatfix]{revtex4}
\usepackage{graphicx,float,amsmath,amssymb}

\newfont{\ensmathquatorze}{msbm10 scaled 1400}
\newfont{\ensmathonze}{msbm10 scaled 1100}
\newfont{\ensmathdix}{msbm10}
\newfont{\ensmathneuf}{msbm10 scaled 833}
\newfont{\ensmathhuit}{msbm10 scaled 694}
\newfam\ensmathfam                        
\textfont\ensmathfam=\ensmathonze        
\scriptfont\ensmathfam=\ensmathdix       
\scriptscriptfont\ensmathfam=\ensmathhuit
\def\ensmf{\fam\ensmathfam\ensmathonze}         

\renewcommand{\leq}{\leqslant}

\def\eqlaw{\stackrel{\mbox{\tiny\rm (law)}}{=}}     
\newcommand{\ket}[1]{|\kern.3ex#1\kern.3ex\rangle}
\newcommand{\bra}[1]{\langle\kern.3ex #1 \kern.3ex|}

\newcommand{\PROPTO}[1]{                
   {{\raisebox{-.3cm}{$\textstyle\propto$}} \atop {\scriptstyle{#1}}}}
\newcommand{\mean}[1]{\left\langle #1 \right\rangle} 
\newcommand{\smean}[1]{\langle #1 \rangle} 

\newcommand{\EXP}[1]{{\mbox{\large e}}^{#1}}         
\newcommand{\argcosh}{\mathop{\mathrm{argcosh}}\nolimits}

\newcommand{\re}{\mathop{\mathrm{Re}}\nolimits}      
\newcommand{\sign}{\mathop{\mathrm{sign}}\nolimits}  
\renewcommand{\min}[2]{\mathop{\mathrm{min}}\nolimits\left( #1 , #2\right)}  
\renewcommand{\max}[2]{\mathop{\mathrm{max}}\nolimits\left( #1 , #2\right)}

\def\RR{{\ensmf R}}                 

\def\I{{\rm i}}                  
\def\D{{\rm d}}                  
\def\Dc{{\rm D}}                 

\newcommand{\sw}{S}

\begin{document}

\title{Dephasing due to electron-electron interaction in a diffusive ring}

\author{Christophe Texier}
\affiliation{Laboratoire de Physique Th\'eorique et Mod\`eles Statistiques,
              UMR 8626 du CNRS, Universit\'e Paris-Sud, 91405 Orsay, 
              France.}
\affiliation{Laboratoire de Physique des Solides, UMR 8502 du CNRS,
              Universit\'e Paris-Sud, 91405 Orsay, France.}

\author{Gilles Montambaux}
\affiliation{Laboratoire de Physique des Solides, UMR 8502 du CNRS,
              Universit\'e Paris-Sud, 91405 Orsay, France.}

\date{October 20, 2006}

\begin{abstract}
  We study the effect of the electron-electron interaction on the weak
  localization correction of a ring pierced by a magnetic flux.  We compute
  exactly the path integral giving the magnetoconductivity for an isolated
  ring.  The results are interpreted in a time representation. This allows
  to characterize the nature of the phase coherence relaxation in the ring.
  The nature of the relaxation depends on the time regime (diffusive or
  ergodic) but also on the harmonics $n$ of the magnetoconductivity.  Whereas
  phase coherence relaxation is non exponential for the harmonic $n=0$, it is
  always exponential for harmonics $n\neq0$.  Then we consider the case of a
  ring connected to reservoirs and discuss the effect of connecting wires. We
  recover the behaviour of the harmonics predicted recently by Ludwig \& Mirlin
  [Phys.~Rev.~{\bf B69} (2004) 193306]
  for a large perimeter (compared to the Nyquist length). We also predict a
  new behaviour when the Nyquist length exceeds the perimeter.
\end{abstract}




\maketitle


\section{Introduction}

In the classical description of transport in weakly disordered metals, elastic
scattering by impurities leads to the finite Drude conductivity at low
temperature. It is well-known that quantum interferences manifest themselves
through a small sample dependent contribution, whose average, denoted by
$\smean{\Delta\sigma}$, is called the weak localization correction.  Dephasing
strongly affects weak localization, which provides a powerful tool to probe
phase coherence in disordered metals. The simplest approach to describe
dephasing is to assume that the time dependence of the phase coherence
relaxation is exponential. Such relaxation can be due for example to magnetic
impurities \cite{HikLarNag80}. It is characterized by a time scale called the
phase coherence time $\tau_\varphi$ and the weak localization correction to
the conductivity takes the form
\begin{eqnarray}\label{sigma1} 
  \smean{\Delta\sigma} =
  -2 \frac{e^2 D}{\pi} \int_0^\infty \D t\, 
  {\cal P}(t)\, \EXP{- t/\tau_\varphi} 
  \:,
\end{eqnarray}
where the so-called cooperon ${\cal P}(t)$ is the contribution to the return
probability originating from quantum interferences between time reversed
trajectories. It is solution of a diffusion equation.  The factor $2$ stands
for spin degeneracy and $D$ is the diffusion coefficient. We have set
$\hbar=1$.  In a quasi-one-dimensional infinite wire, the probability is well
known to vary as ${\cal P}(t)=1/S\sqrt{4 \pi D t}$, where $S$ is the cross
section of the wire, so that the weak localization correction has the familiar
form
\begin{eqnarray} \label{sigma2} 
   \smean{\Delta\sigma} = - \frac{e^2}{\pi\sw} L_\varphi 
\end{eqnarray}
where $L_\varphi= \sqrt{D \tau_\varphi}$ is called the phase
coherence length.


The measurement of the weak localization correction is possible thanks to its
sensitivity to an external magnetic field.  
For a wire \cite{AltAro81}, the effect of a weak perpendicular magnetic
field can be described by introducing an exponential reduction factor
$\EXP{-t/\tau_B}$ in eq.~(\ref{sigma1}), where the characteristic time is
$\tau_B=3/(De^2B^2S)$ (for a wire of square cross section). Consequently the
weak localization is given by eq.~(\ref{sigma2}) with the addition of the
inverse times, ``\`a la Matthiesen''~:
\begin{eqnarray} \label{sigma22} 
  \smean{\Delta\sigma} = - \frac{e^2 \sqrt{D}}{\pi\sw} 
  \left( \frac{1}{\tau_\varphi} + \frac{1}{\tau_B} \right)^{-1/2} 
  \:.
\end{eqnarray}
From the experimental point of view, this effect is of primary importance,
since the magnetic field acts as a probe in order to study phase coherence
and to extract $\tau_\varphi$ and its temperature dependence.


In the more complicated geometry of a ring, the field is also responsible for
magnetoconductivity oscillations as predicted by Altshuler, Aronov \& Spivak
(AAS) \cite{AltAroSpi81}.  The phase coherent return probability is sensitive
to the flux $\phi$ through the ring. It has the simple harmonics expansion~:
\begin{eqnarray}\label{oscicoop}
  {\cal P}(t) = \frac{1}{S\sqrt{4 \pi D t}} 
  \sum_{n=-\infty}^\infty \EXP{-(nL)^2 /4 D t} \,\EXP{\I n\theta}
\end{eqnarray}
where $\theta=4\pi\phi/\phi_0$ is the reduced flux ($\phi_0=h/e$ is the flux
quantum). Each harmonic corresponds to a number of windings of the diffusive
trajectories around the ring. The relation (\ref{sigma1}), with
(\ref{oscicoop}), immediately leads to the familiar result of AAS for the weak
localization correction to the average conductivity in a ring~:
\begin{eqnarray}\label{AAS0}
  \smean{\Delta\sigma(\theta)} 
  =- \frac{e^2}{\pi\sw}\,L_\varphi\,
  \frac{\sinh(L/L_\varphi)}{\cosh(L/L_\varphi)-\cos\theta}
  \:.
\end{eqnarray}
The harmonics of this $\phi_0/2$-periodic correction decay exponentially 
with the perimeter $L$ of the ring~:
\begin{eqnarray}\label{AAS}
  \smean{\Delta\sigma_n} 
  = \int_0^{2\pi}\frac{\D\theta}{2\pi}\,\smean{\Delta\sigma(\theta)}\,
    \EXP{-\I n\theta}
  = - \frac{e^2}{\pi\sw}\,L_\varphi 
  \,\EXP{- |n|L /L_\varphi}
  .
\end{eqnarray}
The combination of the two effects of the magnetic field, AAS
oscillations and penetration in the wires, is obtained by perfoming the
substitution~: $1/\tau_\varphi\rightarrow1/\tau_\varphi+1/\tau_B$ in
eq.~(\ref{AAS0}).

\vspace{0.25cm}

Despite the exponential damping in eq.~(\ref{sigma1}) describes correctly
several dephasing mechanisms like spin-orbit scattering and spin-flip
\cite{HikLarNag80}, or the effect of an external magnetic field
\cite{AltAro81}, a precise description of the electron-electron interaction
requires a more elaborate treatment. In a pioneering paper, Altshuler, Aronov
\& Khmelnitskii (AAK) \cite{AltAroKhm82,AltAro85} have shown that the
dephasing due to the electron-electron interaction can be described in a
one-particle picture by coupling the electron to a fluctuating classical
electromagnetic field. They obtained a result which can be cast in the form
\begin{eqnarray} \label{damping}
  \smean{\Delta\sigma} = -2 \frac{e^2 D}{\pi}  
  \int_0^\infty
  \D t\, {\cal P}(t)\, f(t/\tau_N)\, \EXP{- t/\tau_\varphi } 
  \:,
\end{eqnarray}
where $f(x)$ is a decreasing dimensionless function.
We have also incorporated an exponential relaxation (if it is due to the
external magnetic field $B$ we have simply $\tau_\varphi\to\tau_B$). 
The characteristic time, called the Nyquist time, is given by~\cite{footnote1}
\begin{eqnarray}\label{Nyquist}
  \tau_N = \left(\frac{\hbar^2\sigma_0 S}{e^2 k_BT \sqrt{D}}\right)^{2/3}
  \:, 
\end{eqnarray}
where $T$ is the temperature and $k_B$ the Boltzmann's constant (in the
following we will set $\hbar=k_B=1$). $\sigma_0=2e^2\rho_0D$ is the Drude
conductivity and $\rho_0$ the density of states at Fermi energy, per spin
channel.  The $T^{-2/3}$ power law has been observed in a variety of
experiments (see for example
refs.~\cite{EchGerBozBogNil93b,GouPiePotEstBir00}) and is the signature of the
electron-electron interaction in quasi-1d wires. In the case of an infinite
wire, AAK found that the weak localization correction is given
by~\cite{AltAroKhm82,footnote1}
\begin{eqnarray} \label{AAKAiry} 
  \smean{\Delta\sigma} = 
  \frac{e^2}{\pi\sw} L_N\,
  \frac{{\rm Ai}(\tau_N/\tau_\varphi)}{ {\rm Ai}'(\tau_N/\tau_\varphi)} 
  \:,
\end{eqnarray}
where ${\rm Ai}(z)$ is the Airy function and ${\rm Ai}'(z)$ its derivative. We
have introduced the Nyquist length $L_N=\sqrt{D\tau_N}$ which characterizes
the scale over which the electron-electron interaction is effective. It can be
conveniently rewritten~as~:
\begin{eqnarray}
  L_N=\left(\frac{\sigma_0 \sw D}{e^2T}\right)^{1/3}
  =\left(\frac{\alpha_d}{\pi}N_c\ell_e L_T^2\right)^{1/3}
  \:,
\end{eqnarray} 
where we have introduced the thermal length
$L_T=\sqrt{D/T}$, the number of channels $N_c$, the elastic mean free path
$\ell_e$ and a numerical factor $\alpha_d$ that depends on the dimension $d$
($\alpha_1=2$, $\alpha_2=\pi/2$ and $\alpha_3=4/3$) 
\cite{AkkMon04,TexMon04,TexMonAkk05}. 
We have expressed the Drude conductivity as
$\sigma_0=2\frac{e^2}{h}{\alpha_dN_c\ell_e}/{\sw}$ 
(the factor 2 stands for spin degeneracy).

\vspace{0.25cm}

In addition to the prediction of the power law $\tau_N\propto T^{-2/3}$ for
the coherence time, an important outcome of the AAK theory is that the result
(\ref{AAKAiry}) obviously breaks the addition rule of inverse times. This
indicates that the phase relaxation characterized by the function
$f(t/\tau_N)$ is indeed non exponential. This function has been calculated
recently in ref.~\cite{MonAkk05} where it was found that it varies 
as~\cite{footnote2}
$\EXP{-(\sqrt{\pi}/4)(t/\tau_N)^{3/2}}$ for $t\lesssim\tau_N$.  However the
study of the function $f(t/\tau_N)$ shows that it is very close to an
exponential $\EXP{-t/2\tau_N}$ and eq.~(\ref{AAKAiry}) only deviates from
$
\smean{\Delta\sigma} = - \frac{e^2 \sqrt{D}}{\pi\sw} 
( \frac{1}{2\tau_N} + \frac{1}{\tau_\varphi} )^{-1/2} 
$,
that is eq.~(\ref{sigma22}), by no more than $4\%$
\cite{PieGouAntPotEstBir03,AkkMon04,MonAkk05}.  This explains why it is very
difficult to observe experimentally the functional form (\ref{AAKAiry}) and
most of the magnetoconductance measurements in wires have been analyzed
assuming the form~(\ref{sigma22}).

\vspace{0.25cm}

In the paper of AAK, only simple geometries (like wire and plane) were
considered and it is not clear how the non exponential relaxation of phase
coherence affects the weak localization for a nontrivial geometry.  In a
recent paper, Ludwig \& Mirlin (LM) \cite{LudMir04} have addressed the
question of dephasing due to the electron-electron interaction in a ring.  The
dephasing is then probed by the harmonics of the magnetoconductance
oscillations. These authors found that these harmonics decay with the
perimeter $L$ of the ring in an unexpected way. LM's result can be cast in the
form~\cite{footnote3}
\begin{eqnarray} 
  \label{behaviourLM}
  \smean{\Delta\sigma_n} \propto  \EXP{-|n|(L/L_N)^{3/2}} 
  \:.
\end{eqnarray}
This result is quite interesting, because this nontrivial non exponential
decay of the harmonics leads to an unexpected $\EXP{-nL^{3/2}T^{1/2}}$
temperature behaviour, instead of the incorrect behaviour
$\EXP{-nL/L_N}=\EXP{-nLT^{1/3}}$ naively expected from a simple substitution
$\tau_\varphi\to\tau_N$ in the AAS harmonics (\ref{AAS}).  It also shows that
the geometry of the system may play an important role in the nature of the
dephasing mechanism.


The work of LM was mainly devoted to the study of Aharonov-Bohm (AB)
oscillations in a single ring. The study of AB amplitude, rather than AAS, is
motivated by the lack of disorder averaging in this case. The amplitude
of AB oscillations is given by the harmonics $\smean{\delta\sigma_n^2}$
of the conductivity correlation function 
$\smean{\delta\sigma(B)\,\delta\sigma(B')}$.
As pointed out by LM, these harmonics are expected to be directly related to
the AAS harmonics by the following relation
\begin{eqnarray} 
  \smean{\delta\sigma_n^2} \sim \frac{L_T^2}{L}\smean{\Delta\sigma_n}
  \:,
\end{eqnarray}
where $L_T=\sqrt{D/T}$ is the thermal length. This expression extends the
result of Aleiner \& Blanter \cite{AleBla02} who studied the relation between
conductance fluctuations and weak localization in a wire and a plane when
dephasing is due to the electron-electron interaction. An important
consequence of this relation is that the effect of dephasing on weak
localization and conductance fluctuations is governed by the same length scale
$L_N$. We re-examine this relation and give a more general proof in
appendix~\ref{sec:genAB}.

\vspace{0.25cm}

In our paper, we reconsider the question of weak localization in a ring in
the presence of
electron-electron interaction. Our main goal is to provide a physical picture
as well as a detailed understanding of the results obtained by LM
\cite{LudMir04}. The physical reason for the geometry dependence of the
dephasing can be understood in the following heuristic way.  For a pair of
time reversed trajectories, we denote by $\Phi$ the random phase brought by
the fluctuating electromagnetic field. Average over the Gaussian fluctuations
of the field is denoted by $\smean{\cdots}_V$. Averaging the phase
$\smean{\EXP{\I\Phi}}_V$ produces an exponential damping
$\EXP{-\frac12\smean{\Phi^2}_V}$ responsible for phase coherence relaxation
(this exponential is related to the function $f(t/\tau_N)$ in
eq.~(\ref{damping})). For a quasi-1d system, the typical damping rate
associated to a diffusive trajectory can be written in the form
\begin{eqnarray}
  \label{dephaheur}
  \frac{\D\smean{\Phi^2}_V}{\D t}
  \sim \frac{e^2 T}{\sigma_0\sw}\, r(t) 
  = \frac{1}{\tau_N}\frac{r(t)}{\sqrt{D \tau_N}} 
  \:,
\end{eqnarray}
where $r(t)$ designates the typical distance explored by the diffusive
trajectory over a time scale $t$. As pointed out in ref.~\cite{Fer04},
eq.~(\ref{dephaheur}) can be understood as a local form of the Johnson-Nyquist
theorem~: since the phase and the potential are related by $\dot\Phi=V$,
eq.~(\ref{dephaheur}) measures the potential fluctuations
$\frac{\D}{\D{t}}\smean{\Phi^2}_V=\int\D{t}\smean{V(t)V(0)}_V=2e^2TR_t$, where
$R_t\sim{r(t)/(\sw\sigma_0)}$ is the resistance of a wire of length $r(t)$. It
is clear from eq.~(\ref{dephaheur}) that the dephasing depends on the nature
of the diffusive trajectories. Two regimes can be distinguished~:
\begin{itemize}
\item {\it The diffusive regime.--} In this regime, the boundaries of the
  system play no role and the diffusion follows the behaviour obtained in an
  infinite wire, therefore $r(t)\sim\sqrt{Dt}$, so that
  \begin{eqnarray} 
    \smean{\Phi^2}_V  \propto \left(\frac{t}{\tau_N}\right)^{3/2} 
    \:.
  \end{eqnarray} 
  Therefore we expect the phase relaxation to be non exponential 
  $f(t/\tau_N)\sim\exp{-(t/\tau_N)^{3/2}}$.
  
\item {\it The ergodic regime.--} In this case, the diffusive trajectories
  explore the whole system. This corresponds to time scales larger than the
  Thouless time $\tau_D = L^2/D$.  The characteristic length is given by the
  size of the system (the ring) $r(t)\sim L$ and
  \begin{eqnarray} 
    \langle \Phi^2 \rangle_V  \propto  \frac{\sqrt{\tau_D}}{\tau_N ^{3/2}}\,t 
    = \frac{t}{\tau_c}
    \:.
  \end{eqnarray} 
  We expect the phase relaxation to be exponential, $\exp{-t/\tau_c}$ , 
  where the time scale 
  \begin{eqnarray}
    \label{eq:deftauc}
    \tau_c = \frac{\sigma_0\sw}{e^2 T L} = \frac{\tau_N ^{3/2}}{\tau_D^{1/2}}
  \end{eqnarray}
  is size dependent and is a nontrivial combination of the Thouless time and
  the Nyquist time.
\end{itemize}
It is clear that the function $f(t/\tau_N)$ in eq.~(\ref{damping}) must be
replaced by a more complicated function $f(t/\tau_N,t/\tau_D)$ to account for
finite size effect and describe both regimes. We expect that the winding
around the ring plays an important role and a dephasing of different nature
for trajectories which enclose the ring (corresponding to the $n\neq0$
harmonics of the flux dependence) and for trajectories which do not encircle
the ring (corresponding to the $n=0$ harmonic). For harmonics $n\neq0$, the
trajectories necessarily explore the whole ring and the phase coherence
relaxation is always exponential. This is the origin for the new behaviour
found by LM. For the harmonic $n=0$ we can distinguish two different regimes
corresponding to non exponential and exponential relaxation. In other terms,
the function $f(t/\tau_N,t/\tau_D)$ depends also on the winding number~$n$.

\vspace{0.25cm}

In order to give a firm basis to these arguments, we have reconsidered the
calculation of LM. These authors have studied a symmetric ring connected by
two arms to reservoirs and have derived the weak localization correction
within an instanton approximation for the functional integral describing the
effect of the fluctuating field. The instanton approach is only valid in the
limit of large perimeter of the ring (compared to the length scale
characterizing the electron-electron interaction), that is when $L\gg L_N$
({\it i.e.} $\tau_D\gg\tau_N$).

In our work we have followed a different strategy. We first consider the case
of an isolated ring for which it is possible to compute exactly the path
integral for any value of $L/L_N$. This path integral formulation is
introduced in section~\ref{sec:pathintegral}, and the exact solution is given
in section~\ref{sec:exactsolution}, where a closed expression for the
harmonics of the magnetoconductivity is provided and analyzed in several
regimes.

Our result agrees with the exponential behaviour (\ref{behaviourLM}) found by
LM. However we will point out that LM estimated an incorrect prefactor
(leading to an incorrect temperature dependence). LM's result can be written
as $\smean{\Delta\sigma_n}_{\rm LM} \sim L_N^{9/4}\EXP{-|n|(L/L_N)^{3/2}}$
whereas we will show that the correct result is
$\smean{\Delta\sigma_n}\sim{L_N}\EXP{-|n|(L/L_N)^{3/2}}$. Note that it could
seem at first sight that the difference comes from the fact that our exact
solution stands for an isolated ring whereas LM considered a connected ring,
however we will see that the two situations are closely related. Moreover the
isolated ring leads to consider a path integral exactly similar to the one
studied by LM which allows us to trace back to the origin of LM's incorrect
prefactor (we have also estimated in appendix~\ref{sec:instanton} the path
integral within the instanton approach used by LM. We treat carefully the
prefactor within this approach).

Our exact solution takes also into account the effect of an exponential
relaxation of phase coherence ($\tau_\varphi$).  We will show that both kinds
of dephasing mechanisms ({\it i.e.} $\tau_N$ and $\tau_\varphi$) combine in a
nontrivial way.

In section~\ref{sec:relaxation} we analyze the exact result in a time
representation, that is we study the function $f(t/\tau_N,t/\tau_D)$ that
generalizes the $f(t/\tau_N)$ of eq.~(\ref{damping}). A special emphasis is
put on the difference between harmonics $n=0$ and $n\neq0$~: although the
phase coherence relaxation for the harmonic $n=0$ is either non exponential
(at short times) or exponential (at large times), it is always exponential for
harmonics $n\neq0$. This analysis in time representation is used in the
section~\ref{sec:efconwires} where we discuss the effect of connecting wires~:
in a transport experiment, the ring is necessary connected to wires that can
strongly affect the magnetoconductance. This has been discussed recently for
the case of exponential relaxation of phase coherence \cite{TexMon05}.
Although we do not expect a strong effect of the connecting wires on the
harmonics in the regime $L\gg L_N$, we will show by some simple arguments that
the behaviour (\ref{behaviourLM}) is strongly modified in the other regime
$L\ll L_N$.


\section{Path integral formulation\label{sec:pathintegral}}

We recall the basic ideas of AAK's approach \cite{AltAroKhm82}. In dimension
$d\leq2$ the dephasing is dominated by small energy transfers. The dephasing
for one electron can be modeled through the coupling of the electron with a
classical fluctuating electric potential $V(r,t)$, whose fluctuations are
given by the fluctuation-dissipation theorem. In a Fourier representation, the
correlations are given by 
$\smean{\tilde V\,\tilde V}_{(\vec q,\omega)}=\frac{2e^2T}{\sigma_0 q^2}$. In
a time-space representation it reads
\begin{eqnarray}\label{fdt}
  \mean{V(\vec r,t)\,V(\vec r\,',t')}_V=\frac{2e^2}{\sigma_0}T\,\delta(t-t')\,
  P_d(\vec r,\vec r\,')
  \:,
\end{eqnarray}
where $\mean{\cdots}_V$ designates averaging over the Gaussian fluctuations of
the potential. $P_d(\vec r,\vec r\,')$ is the diffuson, solution of the
equation $-\Delta P_d(\vec r,\vec r\,')=\delta(\vec r-\vec r\,')$ (it is
understood that the zero mode diverging contribution is not taken into account
in the diffuson of eq.~(\ref{fdt}). This point is discussed in
appendix~\ref{sec:gfr}). Eq.~(\ref{fdt}) clearly exhibits the long range
nature of the spatial correlations.

The description of AAK allows to perform a mean-field-like but
non-perturbative treatment of the electron-electron interaction. 
We denote by $\widetilde{\cal P}$ the cooperon that includes the effect
of the electron-electron interaction. It can
be conveniently written with a path integration over the diffusive
trajectories of the electron
\begin{eqnarray}\label{spAAK}
\widetilde{\cal P}(\vec r,\vec r;t) = 
\mean{
  \int_{\vec r(0)=\vec r}^{\vec r(t)=\vec r}{\cal D}\vec r(\tau)\,
  \EXP{ - \int_0^t\D\tau\,\frac{\dot{\vec r}(\tau)^2}{4D}
        \  +\  \I\Phi[\vec r(\tau)] }
     }_V
\end{eqnarray}
where the phase 
\begin{eqnarray}\label{phase}
  \Phi[\vec r(\tau)] = \int_0^t\D\tau\,
  \left[V(\vec r(\tau),\tau)-V(\vec r(\tau),\bar\tau)\right]
  \:,
\end{eqnarray}
with $\bar\tau=t-\tau$, is the phase difference between the two time reversed
trajectories. In the presence of a static magnetic field, a coupling 
$2\I{e}\int_0^t\D\tau\,\dot{\vec r}(\tau)\cdot\vec A(\vec r(\tau))$ to the
vector potential $\vec A(\vec r)$ must be added in the action. In a ring of
perimeter $L$ pierced by a magnetic flux $\phi$, we have $2eA=\theta/L$, where
$\theta=4\pi\phi/\phi_0$ is the reduced flux.

The cooperon has the structure 
$
\widetilde{\cal P}(\vec r,\vec r;t)=
{\cal P}(\vec r,\vec r;t)\,\smean{\EXP{\I\Phi}}_{V,{\cal C}}
$
where ${\cal P}(\vec r,\vec r;t)$ is the cooperon in the absence of the 
electron-electron interaction and $\smean{\cdots}_{{\cal C}}$ denotes 
averaging over closed Brownian curves.

The average over the Gaussian fluctuations of the field in (\ref{spAAK}) 
can be performed thanks to the relation 
$\smean{\EXP{\I\Phi}}_V=\EXP{-\frac12\smean{\Phi^2}_V}$. 
The term that appears in the action is 
\begin{eqnarray}\label{phaseVa}
  \frac12\smean{\Phi^2}_V
  =\frac{2e^2T}{\sigma_0}\int_0^t\D\tau\, W(\vec r(\tau),\vec r(\bar\tau))
\end{eqnarray}
where we have introduced the function $W$, defined in the most symmetric way
by~:
\begin{eqnarray}
  \label{defW}
  W(\vec r,\vec r\,') = \frac{P_d(\vec r,\vec r)+P_d(\vec r\,',\vec r\,')}{2} 
  - P_d(\vec r,\vec r\,')
  \:.
\end{eqnarray}
For one-dimensional wires of section $\sw$ we have 
$P_d(\vec r,\vec r\,')\to\frac1{\sw}P_d(x,x')$, where $P_d(x,x')$ is now the
one dimensional diffuson. Similar subsitutions holds for $W$ and
$\widetilde{\cal P}$. The cooperon finally reads
\begin{eqnarray}\label{spAAK2}
  \widetilde{\cal P}(x,x;t) &=&
  \int_{x(0)=x}^{x(t)=x}{\cal D}x(\tau)\,
  \\\nonumber
  &&
  \times\EXP{ -\int_0^t\D\tau\,
        \Big[\frac{\dot x(\tau)^2}{4D} 
       + \frac{2}{\sqrt{D}\tau_N^{3/2}}\,W(x(\tau),x(\bar\tau))\Big] }
  \:,
\end{eqnarray}
where we have used the definition of the Nyquist time (\ref{Nyquist}).
The weak localization correction is now given by 
\begin{eqnarray}
  \label{spAAK2:cond}
   \smean{\Delta\sigma} =
  -2 \frac{e^2 D}{\pi\sw} \int_0^\infty \D t\, 
  \widetilde{\cal P}(x,x;t)\, \EXP{- t/\tau_\varphi}  
  \:,
\end{eqnarray}
which is another way to write eq.~(\ref{damping}) (note that for a 
translation device such as a wire or a ring,  $\widetilde{\cal P}(x,x;t)$ is 
independent on $x$).

\vspace{0.25cm}

\noindent{\bf Scaling.--}
If we introduce the dimensionless variables $u=\tau/t$ and $y=x/\sqrt{Dt}$,
using the expression of $W(x,x')$ given below by eq.~(\ref{fctW}), we get, for
a ring or for a finite connected wire~:
\begin{eqnarray}
  \label{eq:scaling}
  &&\hspace{-0.25cm}
  \widetilde{\cal P}(x,x;t) = \frac{1}{\sqrt{Dt}}
  \int_{y(0)=y}^{y(1)=y}{\cal D}y(u)\,
  \\\nonumber
  &&\hspace{-0.25cm}
  \EXP{ -\int_0^1\D u\,
        \Big[\frac{1}{4} \dot y(u)^2
       + 2(\frac{t}{\tau_N})^{3/2}
          |y(u)-y(\bar u)|
          \left(1-(\frac{t}{\tau_D})^{1/2}|y(u)-y(\bar u)|\right)\Big] }
  \:,
\end{eqnarray}
where $\bar u=1-u$. We obtain the structure assumed in the introduction~:
\begin{eqnarray}
  \widetilde{\cal P}(x,x;t) = {\cal P}(x,x;t)\times
  f\!\left( \frac{t}{\tau_N},\frac{t}{\tau_D} \right)
  \:,
\end{eqnarray}
where $f(x,y)$ is a dimensionless function. 
Finally it is clear that the integration over time, eq.~(\ref{spAAK2:cond}),  
leads to a conductivity of the form
\begin{eqnarray}
  \smean{\Delta\sigma}=\frac{e^2}{\sw}L\times
  g\!\left( \frac{L}{L_N},\frac{L}{L_\varphi}\right)
  \:,
\end{eqnarray}
where $g(x,y)$ is a dimensionless function.

In the following we will omit the section of the wire.

\vspace{0.25cm}

\noindent{\bf How to get rid of time-nonlocality~?--}
The expressions (\ref{spAAK},\ref{spAAK2}) were the starting point of
ref.~\cite{AltAroKhm82} in which the correction was computed in the case of an
infinite plane and an infinite wire. A first difficulty to evaluate the path
integral (\ref{spAAK2}) is the nonlocality in time of the action. This problem
can be overcome thanks to the translation invariance which makes the function
$W(x,x')$ a function of the difference $x-x'$ only. Such a property is true
only in few cases. More precisely, for the infinite wire (AAK) and a finite
isolated wire, the function is given by $W(x,x')=\frac12|x-x'|$. For the
connected wire and the isolated ring, it reads (see appendix~\ref{sec:gfr})
\begin{eqnarray}\label{fctW}
  W(x,x')=\frac12|x-x'|\left(1-\frac{|x-x'|}{L}\right)
  \:.
\end{eqnarray}

For a translation invariant problem we can follow the strategy of AAK~:
separate the path integral into two parts over the time intervals $[0,t/2]$
and $[t/2,t]$ with
$
\int_{x,0}^{x,t}{\cal D}x(\tau)\rightarrow
\int\D x'\int_{x',t/2}^{x,t}{\cal D}x_1(\tau)
\int_{x,0}^{x',t/2}{\cal D}x_2(\tau)
$,
then perform the change of variables 
$R(\tau)=\frac{x_1(\tau)+x_2(\bar\tau)}{\sqrt2}$ and 
$\rho(\tau)=\frac{x_1(\tau)-x_2(\bar\tau)}{\sqrt2}$ (the Jacobian is 1). 
Since $W(x,x')$ is only function of $x-x'$, the action possesses a potential
term $W(x_1(\tau),x_2(\bar\tau))=W(\rho(\tau),0)$ function of $\rho$ only, 
therefore local in time.

This result is actually due to a general property mentioned in 
ref.~\cite{ComDesTex05}. 
The path integral (\ref{spAAK2}) extends over all Brownian paths coming 
back to their initial value (these paths are called Brownian bridges).
If we consider a Brownian bridge $(x(\tau),0\leq\tau\leq t\,|\,x(0)=x(t)=0)$
we can write the following equality in law~:
\begin{eqnarray}\label{anicerel}
  x(\tau)-x(t-\tau) \eqlaw x(2\tau) 
  \hspace{0.25cm} \mbox{ for } 
  \tau\in[0,t/2]
  \:.
\end{eqnarray}
(Two random variables distributed according to the same 
probability distribution are said to be ``equal in law'').
Therefore it follows that 
$
\int_0^t\D\tau\,{\cal V}(x(\tau)-x(\bar\tau))
\eqlaw\int_0^t\D\tau\,{\cal V}(x(\tau))
$
for any function ${\cal V}(x)$. This immediately gives
\begin{eqnarray}\label{getridnonloc}
  \int_{x(0)=x}^{x(t)=x}{\cal D}x(\tau)\,
    \EXP{ -\int_0^t\D\tau\,
          \left[
            \frac{\dot x(\tau)^2}{4D}+{\cal V}(x(\tau)-x(\bar\tau))
          \right] }
  \nonumber\\
  =\int_{x(0)=0}^{x(t)=0}{\cal D}x(\tau)\,
    \EXP{ -\int_0^t\D\tau\,
          \left[\frac{\dot x(\tau)^2}{4D}+{\cal V}(x(\tau))\right] }
  \:.
\end{eqnarray}


\section{Exact calculation of the path integral in the isolated ring\label{sec:exactsolution}}

From eqs.~(\ref{spAAK2},\ref{getridnonloc}) we can write the weak localization
in the form~:
\begin{eqnarray}
  \smean{\Delta\sigma}
  &=& -2\frac{e^2}{\pi}
  \int_0^\infty\D t\,\EXP{-t/L_\varphi^2}
    \int_{x(0)=0}^{x(t)=0}{\cal D}x(\tau)\,\nonumber\\
  && \times \EXP{ \int_0^t\D\tau\,
          \left[ - \frac{\dot x^2}{4} + 2\I e\dot x\,A(x)
                 - \frac{2}{L_N^3}\,W(x(\tau),0)\right] } \\
  \label{eqjesaispasquoi}
  && \hspace{-2cm}=  -\frac{2e^2}{\pi}\:
  \bra{x=0}
    \frac1{
      \frac1{L_\varphi^2} - \Dc_x^2
      +\frac{1}{L_N^3}\left(|x| - \frac{1}{L}x^2\right)
    }
  \ket{x=0}
\end{eqnarray}
where we have rescaled the time to get rid of the diffusion constant, and
introduced the coupling to the magnetic field. Inside the ring the vector
potential is given by $2eA(x)=\theta/L$, therefore the covariant derivative is
$\Dc_x=\D/\D{x}-\I\theta/L$.

\mathversion{bold}
\subsection{The result for $L_\varphi=\infty$}
\mathversion{normal}

As we have mentioned in the introduction, the most striking effect of the
electron-electron interaction is a modification of the dependence of the AAS
oscillations as a function of $L/L_N$. For this reason we begin this section
by emphasizing the results for two limiting cases demonstrated below.

\vspace{0.25cm}

\noindent$\bullet$ Small perimeter $L\ll L_N$.
In this case~:
\begin{eqnarray}\label{harmpetitL}
  \smean{\Delta\sigma_n} \simeq -\frac{e^2}{\pi}\,{L}\,
  \sqrt6 \left(\frac{L_N}{L}\right)^{3/2}
  \EXP{-|n|\frac{1}{\sqrt6} (\frac{L}{L_N})^{3/2}}
  \:.
\end{eqnarray}

\vspace{0.25cm}

\noindent$\bullet$ Large perimeter $L\gg L_N$.
This regime is the one studied by  LM \cite{LudMir04,footnote3}. We 
obtain
\begin{eqnarray}\label{harmgrandL}
  \smean{\Delta\sigma_n} \simeq \frac{e^2}{\pi}\,{L_N}\,
  \frac{{\rm Ai}(0)}{{\rm Ai}'(0)}\:
  \left(\frac{\sqrt3}2\right)^{|n|}\EXP{-|n|\frac\pi8 (\frac{L}{L_N})^{3/2}}
  \:,
\end{eqnarray}
where 
$
\frac{{\rm Ai}(0)}{{\rm Ai}'(0)}
=-\frac{\Gamma(1/3)}{3^{1/3}\Gamma(2/3)}\simeq-1.372
$.  
The exponential behaviour coincides exactly with the one obtained by LM
(see~\cite{footnote1}). However our prefactor differs as will be discussed in
section~\ref{sec:efconwires}.

\subsection{Computation of the cooperon}

Starting from eq.~(\ref{eqjesaispasquoi}) we introduce the rescaled variable
$\chi=x/L$. Then the weak localization correction rewrites
\begin{eqnarray}
  \smean{\Delta\sigma} = -2\frac{e^2}{\pi}{L}\:C(0,0)
  \:,
\end{eqnarray}
where the Green's function is solution of
\begin{eqnarray}\label{eqcooperon}
  \left[
    -\left(\frac{\D}{\D\chi}-\I\theta\right)^2
    +\frac{L^3}{L_N^3}\left(|\chi|-\chi^2\right)
    +\frac{L^2}{L_\varphi^2}
  \right]C(\chi,\chi')\nonumber\\ = \delta(\chi-\chi')
\end{eqnarray}
for $\chi\in[0,1]$ and periodic boundary conditions.
We introduce the notation
\begin{eqnarray}
  a=\left(\frac{L}{L_N}\right)^3
  \hspace{0.25cm}\mbox{ and }\hspace{0.25cm}
  b=\left(\frac{L}{L_\varphi}\right)^2
  \:.
\end{eqnarray}
We first consider the Cauchy problem. The differential equation
\begin{eqnarray}\label{thediffeq}
  \left[-\frac{\D^2}{\D\chi^2} + a\,\chi(1-\chi)+b\right]f(\chi)=0
\end{eqnarray}
is a hypergeometric equation and can be solved by standard methods
\cite{NikOuv83} which suggest the following transformation
$f(\chi)=\EXP{-\frac12 s^2}\,y(s)$, where the variable is
\begin{eqnarray}
  s=\EXP{\I\frac\pi4}a^{1/4}\left(\chi-\frac12\right) 
  \:.
\end{eqnarray}
It is now easy to see that the function $y(s)$ is solution of the Hermite 
equation. A solution of (\ref{thediffeq}) is
\begin{eqnarray}\label{anicesol}
  \tilde f(\chi) = \EXP{-\frac12 s^2}\, H_\nu(s)
  \:,
\end{eqnarray}
where $H_{\nu}(s)$ is the Hermite function (see appendix~\ref{sec:Hermite}). 
The index $\nu$ reads
\begin{eqnarray}
  \nu=-\frac12+\I\omega
  \hspace{0.5cm}\mbox{with}\hspace{0.5cm}
  \omega=\frac{\sqrt{a}}{8} +  \frac{b}{2\sqrt{a}}
  \:.
\end{eqnarray}
Thanks to the symmetry of the differential equation with respect to the 
substitution $\chi\leftrightarrow1-\chi$, and since $\tilde f(\chi)$ is not
invariant under this transformation, another possible solution is
$\tilde f(1-\chi)=\EXP{-\frac12 s^2}\,H_{\nu}(-s)$.

In order to construct the Green's function $C(\chi,\chi')$ we introduce 
another solution of eq.~(\ref{thediffeq}) satisfying the boundary conditions 
\begin{eqnarray}\label{boundary}
  f(0)=1 \hspace{0.25cm}\mbox{and}\hspace{0.25cm} f(1)=0
  \:.
\end{eqnarray}
This solution is related to $\tilde f(\chi)$ by
\begin{eqnarray}\label{relfft}
  f(\chi) = \frac{\tilde f(0)\tilde f(\chi)-\tilde f(1)\tilde f(1-\chi)}
                 {\tilde f(0)^2-\tilde f(1)^2}
  \:.
\end{eqnarray}
The derivatives of $f(\chi)$ at $\chi=0$ and $1$ will be central quantities 
in the  following. They can be related to the derivatives of $\tilde f(\chi)$ 
as
\begin{eqnarray}
  f'(0) &=& \frac{\tilde f(0)\tilde f'(0)+\tilde f(1)\tilde f'(1)}
                 {\tilde f(0)^2-\tilde f(1)^2}
  \\
  f'(1) &=& \frac{\tilde f(0)\tilde f'(1)+\tilde f(1)\tilde f'(0)}
                 {\tilde f(0)^2-\tilde f(1)^2}
  \:.
\end{eqnarray} 
Then the cooperon is given by \cite{Des00,TexMonAkk05}~:
\begin{eqnarray}\label{expressionC}
  C(\chi,\chi')&=&\frac{\EXP{\I\theta(\chi-\chi')}}{{\cal M}}
  \Big[
     f(\chi)\,f(\chi') + \EXP{\I\theta}\,f(\chi)\,f(1-\chi') 
     \nonumber\\
     &&\hspace{-0.5cm}
     +\EXP{-\I\theta}\,f(1-\chi)\,f(\chi') + f(1-\chi)\,f(1-\chi') 
  \Big]\nonumber\\
  &&\hspace{-1.5cm}
  -\frac{\EXP{\I\theta(\chi-\chi')}}{f'(1)}
  f(\max{\chi}{\chi'})\,f(1-\min{\chi}{\chi'})
  \:,
\end{eqnarray}
where ${\cal M}=-2f'(0)+2\cos\theta\,f'(1)$.
At the origin we obtain $C(0,0) = \frac{1}{{\cal M}}$, therefore
\begin{eqnarray}\label{RES1}
  \smean{\Delta\sigma} 
  = -\frac{e^2}{\pi}{L}\:\frac{1}{-f'(0)+f'(1)\,\cos\theta}
  \:.
\end{eqnarray}
This result shows that in the most general case with exponential relaxation
($\tau_\varphi$) and electron-electron interaction ($\tau_N$), the flux
dependence of the weak localization correction has still the same structure as
AAS, eq.~(\ref{AAS0}). As a consequence the harmonics still decay
exponentially with~$n$~:
\begin{eqnarray}
  \smean{\Delta\sigma_n}
  \label{RES3}
  &=& -\frac{e^2}{\pi}{L}\:
  \frac{\EXP{-|n|\ell_{\rm eff}}}{\sqrt{f'(0)^2-f'(1)^2}}
  \:,
\end{eqnarray}
where, by analogy with eq.~(\ref{AAS}), we have introduced an effective
``perimeter'' $\ell_{\rm eff}$, defined by
\begin{eqnarray}\label{RES2}
  \cosh\ell_{\rm eff}=\frac{f'(0)}{f'(1)}
  \:.
\end{eqnarray}
The equations (\ref{RES3},\ref{RES2}) are central results of the 
paper.
We now analyze several limiting cases, which requires a detailed study
of $\tilde f(0)$, $\tilde f'(0)$, $\tilde f(1)$ and $\tilde f'(1)$.

\mathversion{bold}
\subsection{No electron-electron interaction~: $L_N=\infty$.}
\mathversion{normal}

We check first that we recover the result (\ref{AAS}) of AAS. In terms of the
parameters $a$ and $b$, the limit $L_N\to\infty$ corresponds to $a\to0$ with
$b$ finite, then $\omega\simeq b/(2\sqrt a)\to\infty$. With the help of the
asymptotic behaviour (\ref{asy1}) we get $\tilde
f(\chi)\propto\EXP{-\sqrt{b}\,\chi}$ which gives
$f(\chi)=\frac{\sinh\sqrt{b}(1-\chi)}{\sinh\sqrt{b}}$. The derivatives are
$f'(0)=-\sqrt{b}\coth\sqrt{b}$ and $f'(1)=-{\sqrt{b}}/{\sinh\sqrt{b}}$ and
the effective perimeter reads $\ell_{\rm eff}=\sqrt{b}=L/L_\varphi$. The
harmonics are given by eq.~(\ref{AAS}).

\mathversion{bold}
\subsection{Small perimeter $L\ll L_N$.\label{sec:subsecSP}}
\mathversion{normal}

Instead of expanding the exact solution given above for $a\ll1$,
we go back to the differential equation (\ref{thediffeq}) and construct
the solution $f(\chi)$ by a  perturbative approach for the 
small parameter $a$.
This perturbative method is explained in the appendix~\ref{sec:pad}.
It follows from the expressions (\ref{fp0pepe},\ref{fp1pepe}) that
\begin{eqnarray}
  \cosh\ell_{\rm eff} = \cosh\sqrt{b} + \frac{a}{12\sqrt{b}}\sinh\sqrt{b}
  + O(a^2) 
\end{eqnarray}
and 
\begin{eqnarray}
  f'(0)^2-f'(1)^2 
  = b + \frac{a}{2\sqrt{b}}\left(\coth\sqrt{b}-\frac{1}{\sqrt{b}}\right) 
   + O(a^2) 
  \:.
\end{eqnarray}

In the limit $L\ll L_\varphi$, the effective perimeter can be written
\begin{eqnarray}\label{leffsmape}
  \ell_{\rm eff} \simeq 
  \sqrt{\left(\frac{L}{L_\varphi}\right)^2 
  + \frac16\left(\frac{L}{L_N}\right)^3}
  \:.
\end{eqnarray}
This combination is not surprising from eq.~(\ref{thediffeq})
($1/6$ is the average of $\chi(1-\chi)$ over the interval).
The prefactor of the harmonic is given by~:
\begin{eqnarray}\label{prefsmape1}
  \sqrt{f'(0)^2-f'(1)^2} 
  \simeq\ell_{\rm eff}
  \:.
\end{eqnarray}

In the opposite limit $L\gg L_\varphi$, the effective perimeter can be 
expanded as 
\begin{eqnarray}
  \ell_{\rm eff} \simeq
  \frac{L}{L_\varphi}+\frac1{12}\frac{L_\varphi L^2}{L_N^3}+\cdots
\end{eqnarray}
and the prefactor is given by
\begin{eqnarray}\label{prefsmape2}
  \sqrt{f'(0)^2-f'(1)^2}\simeq 
  \frac{L}{L_\varphi} 
  \left(1+\frac14\left(\frac{L_\varphi}{L_N}\right)^3\right) 
  \:.
\end{eqnarray}

For $L_\varphi=\infty$, it is clear that
eqs.~(\ref{leffsmape},\ref{prefsmape1}) with eq.~(\ref{RES3})
give eq.~(\ref{harmpetitL}).

\begin{figure}[!ht]
\begin{center}
\includegraphics[scale=1]{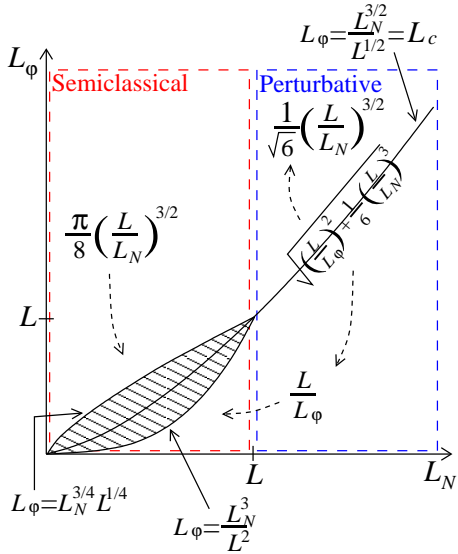}
\end{center}
\caption{ (Color online) 
          {\it On this figure we summarize the different limits for the 
             effective perimeter $\ell_{\rm eff}$ given by eq.~(\ref{RES2}).
             For the regime $L\ll L_N$, we have used a perturbative method 
             in section~\ref{sec:subsecSP}.
             The regime $L\gg L_N$ has been studied with the semiclassical
             approximation (instanton approach) in section~\ref{sec:subsecLP}.
             Since the effective perimeter has the structure~:
             $\ell_{\rm eff}=(L/L_N)^{3/2}\,\eta(L_c^2/L_\varphi^2)$, 
             the line $L_\varphi=L_c\propto L_N^{3/2}$ separates the regimes 
             of large and small $L_\varphi$.
             The dashed area corresponds to the crossover region where 
             the full expression (\ref{lefflarpe}) is needed.
             \label{fig:phasediag} }}
\end{figure}

\mathversion{bold}
\subsection{Large perimeter $L\gg L_N$.\label{sec:subsecLP}}
\mathversion{normal}

In this limit, the function (\ref{anicesol}),
$
\tilde f(\chi)
=\EXP{-\I\frac12\sqrt{a}(\chi-1/2)^2}\,
H_{-\frac12+\I\omega}(\EXP{\I\pi/4}a^{1/4}(\chi-1/2))
$, presents for $\chi\to0$ a behaviour given by (\ref{fH2Ai0}).
For $\chi\to1$ the function is reduced by an exponential factor 
$\EXP{-\pi\omega}$ (in this limit $\omega\simeq\frac{\sqrt{a}}{8}\gg1$). 
It follows from (\ref{relfft}) that
\begin{eqnarray}
  \label{eq55}
  f'(0)\simeq\frac{\tilde f'(0)}{\tilde f(0)}=
  \frac{L}{L_N}\,
  \frac{{\rm Ai}'(L_N^2/L_\varphi^2)}{{\rm Ai}(L_N^2/L_\varphi^2)}
  \:,
\end{eqnarray}
(this expression is also derived in appendix~\ref{sec:instanton} by a 
different method).

\begin{figure}[!ht]
\begin{center}
\includegraphics[scale=0.4]{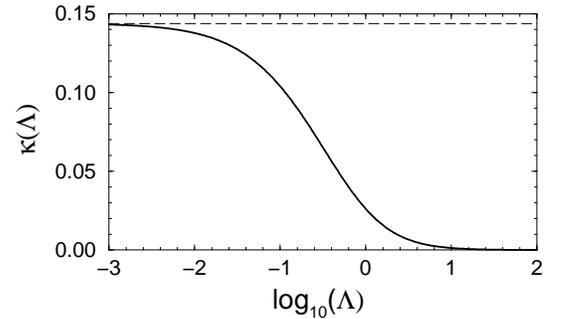}
\end{center}
\caption{\it The function $\kappa(\Lambda)$ of eq.~(\ref{fctkappa}).
         \label{fig:kappa}}
\end{figure}

The relation (\ref{0rel1}) shows that 
$\tilde f'(1)\simeq-\I\EXP{-\pi\omega}\tilde f'(0)$. 
Therefore we expect that $f'(1)/f'(0)\sim\EXP{-\pi\omega}$ which leads 
to $\ell_{\rm eff}\simeq\pi\omega$. 
However this dominant term in $\tilde f'(1)$ is imaginary and does not 
contribute to $f'(1)$ which is given by the next term in the expansion
of $\tilde f'(1)$.
Instead of performing a systematic expansion of $\tilde f'(1)$,
we use the semiclassical solution for the cooperon 
(see appendix~\ref{sec:instanton}) which leads to the behaviour
(\ref{leffinstant},\ref{eq:C25}). As a result 
the effective perimeter is given by the sum of two contributions~:
\begin{eqnarray}\label{lefflarpe}
  \ell_{\rm eff} =
   \left(\frac{L}{L_N}\right)^{3/2}
   \eta\!\left(\frac{L_N^3}{L_\varphi^2L}\right)
  + \kappa(L_N^2/L_\varphi^2)
  \:,
\end{eqnarray}
where
\begin{eqnarray}
  \label{fcteta}
  \eta(x) = 
  \left(\frac{1}{4}+{x}\right)\arctan\frac1{\sqrt{4x}} 
  +\frac{\sqrt{x}}{2}
\end{eqnarray}
(see appendix~\ref{sec:instanton}). The second term involves the small and 
smooth function
\begin{eqnarray}
  \label{fctkappa}
  \kappa(\Lambda) = 
  \ln
  \left( -4\pi\EXP{\frac43\Lambda^{3/2}}
     {\rm Ai}(\Lambda){\rm Ai}'(\Lambda)
  \right)
  \:.
\end{eqnarray}
$\kappa(\Lambda)$ interpolates between $\kappa(\infty)=0$ at large $\Lambda$
and $\kappa(0)=\ln(2/\sqrt3)\simeq0.1438$ at $\Lambda=0$ (see
figure~\ref{fig:kappa}). Since the first term in (\ref{lefflarpe}) is much
larger than 1 for the regime considered in this subsection,
$\kappa(L_N^2/L_\varphi^2)$ can be neglected in most cases.

\vspace{0.25cm}

Finally the weak localization correction reads
\begin{eqnarray}\label{RES4}
  \smean{\Delta\sigma_n} \simeq \frac{e^2}{\pi}\,{L_N}\,
  \frac{{\rm Ai}(L_N^2/L_\varphi^2)}{{\rm Ai}'(L_N^2/L_\varphi^2)}\:
   \EXP{-|n|\ell_{\rm eff} }
  \:.
\end{eqnarray}

We remark that the harmonic $0$ coincides, as it should in the limit $L\gg
L_N$, with the result of AAK, eq.~(\ref{AAKAiry}), for the infinite wire. This
provides another check of the exact solution, and in particular of its 
prefactor.

\vspace{0.25cm} 

We now discuss the various limiting cases obtained by varying $L_\varphi$.
First we remark that the prefactor of eq.~(\ref{RES4}) has the form
$L_N\,g_1(L_N^2/L_\varphi^2)$, where $g_1(x)$ is a dimensionless function,
whereas the effective perimeter has the form 
${\ell}_{\rm eff}=(L/L_N)^{3/2}\,\eta(L_c^2/L_\varphi^2)$ with
$L_c=L_N\sqrt{L_N/L}\ll L_N$ (if we neglect the smooth contribution). 
Therefore we have to distinguish three regimes~:

\vspace{0.25cm}

\noindent$\bullet$ 
{\it Negligible exponential dephasing~: $L_\varphi\gg L_N$.} 
Eqs.~(\ref{lefflarpe},\ref{RES4}) give eq.~(\ref{harmgrandL}).
This is the only regime in which the contribution $\kappa(\Lambda)$ in 
eq.~(\ref{lefflarpe}) plays a role.

\vspace{0.25cm}

\noindent$\bullet$ 
{\it $L_c=L_N\sqrt{L_N/L}\ll L_\varphi\ll L_N$}. 
The prefactor simplifies as
\begin{eqnarray}
  \smean{\Delta\sigma_n}\simeq-\frac{e^2}{\pi}\,{L_\varphi}\,
  \EXP{ -|n|{\ell}_{\rm eff} }
\end{eqnarray}
and the effective perimeter can be expanded as
\begin{eqnarray}
  \ell_{\rm eff} \simeq
    \frac\pi8 \left(\frac{L}{L_N}\right)^{3/2}
  + \frac\pi2 \frac{L_N^{3/2}L^{1/2}}{L_\varphi^2} 
  - \frac43   \left(\frac{L_N}{L_\varphi}\right)^{3/2}
  \:.
\end{eqnarray}
The two first terms correspond to $\pi\omega$.
The effective perimeter is dominated by the first term. However the function
$\eta(b/a)$ appears in the argument of an exponential, in eq.~(\ref{RES4}),
multiplied by a large parameter as $\ell_{\rm eff}=\sqrt{a}\,\eta(b/a)$.
Therefore it is not clear {\it a priori} when the terms of the expansion of
$\eta(x)$ are negligible. 

\vspace{0.25cm}

\noindent$\bullet$ 
{\it Dominant exponential dephasing~: $L_\varphi\ll L_c$}.
In this case the expansion of the effective perimeter reads
\begin{eqnarray}
  \ell_{\rm eff} \simeq \frac{L}{L_\varphi} 
  + \frac1{12}\frac{L_\varphi L^2}{L_N^3}
  \:,
\end{eqnarray}
which coincides with the expansion of the result 
$\ell_{\rm eff}\simeq\sqrt{b+a/6}\simeq\sqrt{b}+\frac{a}{12\sqrt{b}}$ 
obtained by a perturbative expansion in $a$. 
If $L_\varphi\ll L_N(L_N/L)^2$, we recover the result of AAS~(\ref{AAS}).


\hspace{0.25cm}

We have summarized the different limits for the effective perimeter on 
figure~\ref{fig:phasediag}.


\section{Relaxation of phase coherence\label{sec:relaxation}}

In this section we interpret the results of the previous section in a time
representation and give a rigorous presentation of the heuristic discussion of
the introduction. The results of this section may be useful to consider more
complicated situations than an isolated ring, when the path integral cannot be
computed exactly, like the connected ring studied in the next section. Let us
consider the $n$-th harmonic $\widetilde{\cal P}_n(x,x';t)$ of the cooperon.
The Fourier transform over the magnetic flux of the path integrals,
eqs.~(\ref{spAAK},\ref{spAAK2}) in which the coupling to the external magnetic
field has been re-introduced, selects the paths with a winding number equal to
$n$. We can write~:
\begin{eqnarray}
  \widetilde{\cal P}_n(x,x;t)&=&
  \int_0^{2\pi}\frac{\D\theta}{2\pi}\widetilde{\cal P}(x,x;t)\,
  \EXP{-\I n\theta}
  \\ \nonumber
  && \hspace{-2.25cm}=
  \mean{
    \int_{x(0)=x}^{x(t)=x} {\cal D}x(\tau)\,
    \EXP{-\frac1{4}\int_0^t\D\tau\,\dot x(\tau)^2+\I\Phi[x(\tau)]}
    \,\delta_{n,{\cal N}[x(\tau)]}
  }_V
\end{eqnarray}
where ${\cal N}[x(\tau)]$ is the winding number of the trajectory around the
ring.  For a closed trajectory ($x(0)=x(t)$) we have
${\cal{N}}[x(\tau)]=\frac1L\int_0^t\D\tau\,\dot x(\tau)$.  (In this section we
set $D=1$). Let us introduce the probability ${\cal P}_n(x,x';t)$ for a
Brownian curve to go from $x'$ to $x$ in a time $t$ encircling $n$ times the
flux~:
\begin{eqnarray}
  {\cal P}_n(x,x';t) = 
  \int_{x(0)=x'}^{x(t)=x} {\cal D}x(\tau)\,
  \EXP{-\frac1{4}\int_0^t\D\tau\,\dot x(\tau)^2}
  \,\delta_{n,{\cal N}[x(\tau)]}
  \:.
\end{eqnarray}
For an isolated ring this probability simply reads
\begin{eqnarray}
  {\cal P}_n(x,x';t) = \frac{1}{\sqrt{4\pi t}}\,
  \EXP{ -\frac{(x-x'-nL)^2}{4t} }
  \:.
\end{eqnarray}
Then we can rewrite the harmonics of the conductivity as
\begin{eqnarray}\label{eq42}
  \widetilde{\cal P}_n(x,x;t)&=&
  {\cal P}_n(x,x;t)\,
  \mean{
    \EXP{\I\Phi[x(\tau)]}
  }_{V,{\cal C}_n}
  \\
  &=&{\cal P}_n(x,x;t)\,
  \mean{\EXP{-\frac12\smean{\Phi^2}_V}}_{{\cal C}_n}
\end{eqnarray}
with $\smean{\Phi^2}_V$ given by (\ref{phaseVa})~:
\begin{eqnarray}
  \frac12\smean{\Phi^2}_V=\frac2{\tau_N^{3/2}}\int_0^t\D\tau\,
  W(x(\tau),x(\bar\tau))
  \:.
\end{eqnarray}
The average 
\begin{eqnarray}
  \smean{\cdots}_{{\cal C}_n} &=& 
  \frac1{{\cal P}_n(x,x;t)} 
  \\ \nonumber &&\hspace{-1.5cm}\times
  \int_{x(0)=x}^{x(t)=x} {\cal D}x(\tau)\,
  \cdots\,
  \EXP{-\frac1{4}\int_0^t\D\tau\,\dot x(\tau)^2}
  \,\delta_{n,{\cal N}[x(\tau)]}
\end{eqnarray} 
is performed over all closed Brownian trajectories with winding $n$. The
probability ${\cal P}_n$ in the denominator ensures the normalization
$\smean{1}_{{\cal C}_n}=1$.  Therefore the function
$\smean{\EXP{\I\Phi}}_{V,{\cal C}_n}$, related to the $n$-th harmonic of the
AAS oscillations, characterizes the relaxation of phase coherence due to the
electron-electron interaction for trajectories with winding number $n$. We now
analyze this quantity.

\subsection{Diffusion of the phase\label{sec:diffphase}}

An indication on the nature of the phase relaxation, characterized by the
function $\smean{\EXP{-\frac12\smean{\Phi^2}_V}}_{{\cal C}_n}$, can be obtained
by studying the diffusion of the phase, that is the much simpler quantity
$\smean{\Phi^2}_{V,{\cal C}_n}$.  If we consider $\tau<t/2$, the average
$\smean{W}_{{\cal C}_n}$ is given by
\begin{eqnarray}
  &&{\cal P}_n(x,x;t)\mean{W(x(\tau),x(\bar\tau))}_{{\cal C}_n}
  =\frac1L \int_0^L\D x\D x'\, W(x,x')\,
  \nonumber\\
  &&\hspace{1cm}\times\sum_{m=-\infty}^{+\infty} 
  {\cal P}_m(x,x';2\tau)\,{\cal P}_{n-m}(x',x;t-2\tau)
  \:.
\end{eqnarray}
The final result is symmetric with respect to $2\tau\leftrightarrow t-2\tau$. 
The double integration can be reduced to a simple integration thanks to
the relation
$\int_0^1\D x\D x'\,f(x-x')=\int_0^1\D u\,(1-u)[f(u)+f(-u)]$. Then
the integral is unfolded to extend over $\RR$. We obtain~:
\begin{eqnarray}\label{smallphaseanalysis}
  &&{\cal P}_n(x,x;t)\mean{W(x(\tau),x(\bar\tau))}_{{\cal C}_n}
=\int_{-\infty}^{+\infty}\D x\,\Omega(x)\, 
\nonumber\\
  &&\hspace{0.5cm}\times \frac12\left(
    \frac{ \EXP{-\frac{x^2}{8\tau}} }{ \sqrt{8\pi\tau} }
    \frac{ \EXP{-\frac{(x-nL)^2}{4(t-2\tau)}} }{ \sqrt{4\pi(t-2\tau)} }
    + \left( n \rightarrow -n \right)
  \right)
\end{eqnarray}
where $\Omega(x)$ is the function
$\Omega(x)=2W(x,0)(1-\frac{x}{L})=x(1-\frac{x}{L})^2$ for $x\in[0,L]$
and periodised on $\RR$.
To go further we distinguish two cases depending on the relative
order of magnitude of the time $t$ and the Thouless time $\tau_D$.

\mathversion{bold}
\subsubsection*{Diffusive regime ($t\ll\tau_D$)}
\mathversion{normal}

In this case we have to separate the cases $n=0$ and $n\neq0$.

\vspace{0.25cm}

\noindent$\bullet$
\mathversion{bold}
{\bf Harmonic $n=0$.}
\mathversion{normal}
The integral is dominated by the neighbourhood of
$x\sim0$ since the Gaussian function is very narrow compared to $L$. 
We can replace the function $\Omega(x)$ by its behaviour
near the origin~: $\Omega(x)\to\theta(x)x$, where $\theta(x)$ is the 
Heaviside function. 
We obtain
\begin{eqnarray}
  \mean{W(x(\tau),x(\bar\tau))}_{{\cal C}_0}
  \simeq \sqrt{\frac{2\tau(t-2\tau)}{\pi t}}
  \:.
\end{eqnarray}
This result should be symmetrised for $\tau>t/2$.
Integration over time $\tau$ gives 
\begin{eqnarray}\label{phidiff0st}
  \frac12\smean{\Phi^2}_{V,{\cal C}_0}\simeq
  \frac{\sqrt\pi}{4} \left(\frac{t}{\tau_N}\right)^{3/2}
  \:.
\end{eqnarray}
In this regime the geometry plays no role and we recover the result 
obtained for an infinite wire \cite{AkkMon04,MonAkk05}.
This result is related to the AAK behaviour (\ref{AAKAiry}) for 
$L_N\gg L_\varphi$.

\vspace{0.25cm}

\noindent$\bullet$
\mathversion{bold}
{\bf Harmonic $n\neq0$.}
\mathversion{normal}
For times $t\ll\tau_D$,
the integral (\ref{smallphaseanalysis}) can be estimated by
the steepest descent method~:
it is dominated by the neighbourhood of the point minimizing
$\frac{x^2}{8\tau}+\frac{(x-nL)^2}{4(t-2\tau)}$, that is
$x=nL\,2\tau/t$. Therefore we obtain~:
\begin{eqnarray}
  \mean{W(x(\tau),x(\bar\tau))}_{{\cal C}_n}
  \simeq \frac12
  \left[ \Omega(nL\,2\tau/t) + \Omega(-nL\,2\tau/t) \right]
  \:.
\end{eqnarray}
The integration over time leads to average the function $\Omega$~:
$\int_0^t\D t\,\mean{W}_{{\cal C}_n}=t\int_0^L\frac{\D x}{L}\,\Omega(x)$.
It immediately follows that~:
\begin{eqnarray}\label{phidiffnst}
  \frac12\smean{\Phi^2}_{V,{\cal C}_n}\simeq
  \frac{1}{6}\frac{\tau_D^{1/2}}{\tau_N^{3/2}}\,t
  = \frac{1}{6}\frac{t}{\tau_c}
  \:.
\end{eqnarray}

\mathversion{bold}
\subsubsection*{Ergodic regime ($t\gg\tau_D$)}
\mathversion{normal}

The cases $n=0$ and $n\neq0$ can be treated on the same footing.
The Gaussian function in eq.~(\ref{smallphaseanalysis}) is very broad 
compared to $L$ and we can replace $\Omega(x)$ by  its average value
$\Omega(x)\to\frac1L\int_0^L\D x\,\Omega(x)=\frac{L}{12}$. It follows that
$\smean{W}_{{\cal C}_n}\simeq\frac{L}{12}$, then~:
\begin{eqnarray}\label{phidifflt}
  \frac12\smean{\Phi^2}_{V,{\cal C}_n}\simeq
  \frac{1}{6}\frac{t}{\tau_c}
  \hspace{0.75cm} \forall\ n
  \:.
\end{eqnarray}

To summarize we see that, for the harmonic $n=0$, the diffusion of the phase
crosses over from a $t^{3/2}$ behaviour to a linear $t$ behaviour, whereas for
$n\neq0$ it behaves always linearly. This difference shows up in the 
function $\smean{\EXP{\I\Phi}}_{V,{\cal C}_n}$ leading either to a non 
exponential or to an exponential phase coherence relaxation.

\mathversion{bold}
\subsection{The function $\smean{ \EXP{\I\Phi} }_{V,{\cal C}_n}$}
\mathversion{normal}

The calculation of $\smean{ \EXP{\I\Phi} }_{V,{\cal C}_n}$ is a more 
difficult task. It can be obtained by different strategies.

\vspace{0.25cm}

\noindent(A) {\it Small phase approximation.--}
At short times, the phase $\Phi$ is small, therefore we can linearize the
exponential so that~:
$
\smean{\EXP{-\frac12\smean{\Phi^2}_V}}_{{\cal C}_n}
\simeq\EXP{-\frac12\smean{\Phi^2}_{V,{\cal C}_n}}
$.
Then we can use the results given above in subsection~\ref{sec:diffphase}.

\vspace{0.25cm}

\noindent(B) {\it Inverse Laplace transform.--} 
The weak localization correction to the conductivity 
$
\smean{\Delta\sigma_n} \sim \int_0^\infty\D t\,\EXP{-t/\tau_\varphi}\,
{\cal P}_n(x,x;t)\,\smean{\EXP{\I\Phi}}_{V,{\cal C}_n}
$
has been derived for arbitrary $\tau_\varphi=L_\varphi^2$. 
Physically, the parameter
$\tau_\varphi$ takes into account other dephasing mechanisms
responsible for an exponential relaxation of phase coherence. From a technical
point of view the parameter $\tau_\varphi$ allows to probe time scale
$t\sim\tau_\varphi$ in the path integral and we can,  in
principle, compute the inverse Laplace tranform of $\smean{\Delta\sigma_n}$.

\vspace{0.25cm}

\noindent(C) {\it Large phase for $n\neq0$.--}
If none of the previous methods can be used (the method (B) because it is
too difficult, and the method (A) because it is not the range of interest),
we can use the following remark~:
for $\tau_\varphi=\infty$ we see from (\ref{harmpetitL}) that
\begin{eqnarray}
  \smean{\Delta\sigma_n}\sim
  \int_0^\infty\D t\,\frac{1}{\sqrt{t}} \EXP{-\frac{(nL)^2}{4t}}
  \smean{\EXP{\I\Phi}}_{V,{\cal C}_n} 
  \:.
\end{eqnarray}
We expect that the behaviour at large time involves the tail of
$\smean{\EXP{\I\Phi}}_{V,{\cal C}_n}$ which we can assume to behave as
$\smean{\EXP{\I\Phi}}_{V,{\cal C}_n}\propto t^\mu\EXP{-\Upsilon\, t^\alpha}$.
The integral of the {\it l.h.s} can be estimated by the steepest descent
method
\begin{eqnarray}\label{steepdes}
  &&\int_0^\infty\D t\,\frac{1}{\sqrt{t}} \EXP{-\frac{(nL)^2}{4t}}\:
  t^\mu\EXP{-\Upsilon\, t^\alpha}
  \nonumber\\
  &&\simeq
  \sqrt{\frac{2\pi}{\alpha(\alpha+1)\Upsilon}}
  \ t_*^{\mu-\frac\alpha2 +\frac12}
  \:\EXP{-\Upsilon(\alpha+1)t_*^\alpha}
  \:,
\end{eqnarray}
where $t_*=[\frac1{\alpha \Upsilon}(\frac{nL}{2})^2]^{\frac1{\alpha+1}}$. The
coefficient $\Upsilon$ and the exponents $\alpha$ and $\mu$ are obtained by
comparison of the dependence of this result with $L$ and $n$ with the known
behaviour for $\smean{\Delta\sigma_n}$.

\mathversion{bold}
\subsection{Small perimeter $L\ll L_N$}
\mathversion{normal}

We now analyze $\smean{\EXP{\I\Phi}}_{V,{\cal C}_n}$ in the small perimeter
limit.

\subsubsection*{a. Short times}

In the short time limit, the linearization of the exponential is valid (method
(A)).  Therefore we can use expressions
(\ref{phidiff0st},\ref{phidiffnst},\ref{phidifflt}).  These expressions give a
precise definition of the ``short time'' regime, which extends until
$\smean{\Phi^2}_{V,{\cal C}_n}\sim1$ that is $t\sim\tau_c$. The time scale
$\tau_c$, given by eq.~(\ref{eq:deftauc}), is associated to the length scale
$L_c=L_N\sqrt{{L_N}/{L}}$ introduced in subsection~\ref{sec:subsecLP} 
(see~\cite{footnote1}).

\subsubsection*{b. Long times}

In this case we consider the harmonic $n=0$ and $n\neq0$ on the same footing.
The regime $t\gg\tau_D$ corresponds to
$L\ll L_\varphi$. With $L\ll L_N$ this leads to the ``perturbative'' regime 
$a,b\ll1$ for which we have found the 
expressions~(\ref{leffsmape},\ref{prefsmape1})~:
\begin{eqnarray}
  &&\int_0^\infty\D t\,\EXP{-t/\tau_\varphi}\,
  \widetilde{\cal P}_n(x,x;t) 
  \nonumber\\
  &=& 
  \frac1{2\sqrt{\frac1{L_\varphi^2}+\frac16\frac{L}{L_N^3}}}\,
  \EXP{-|n|L\sqrt{\frac1{L_\varphi^2}+\frac16\frac{L}{L_N^3}}}
  \:.
\end{eqnarray}
The inverse Laplace transform can be computed exactly in this case. It gives
\begin{eqnarray}
  \widetilde{\cal P}_n(x,x;t) 
  =\frac{1}{2\sqrt{\pi t}}\,\EXP{-\frac{(nL)^2}{4t}}\,\EXP{-\frac{L}{6L_N^3}t}
  \:.
\end{eqnarray}
We immediately obtain
\begin{eqnarray}
  \mean{\EXP{\I\Phi}}_{V,{\cal C}_n} = 
  \exp{-\frac{1}{6} \frac{t}{\tau_c} }
  \hspace{0.5cm}\mbox{ for } t\gg\tau_D\ \forall n
  \:.
\end{eqnarray}
This is the same result as for $t\ll\tau_c$. Since $\tau_D\ll\tau_c$, it turns 
out that the result obtained from the linearization of the exponential is 
valid for all times.

\subsubsection*{c. Summary}

%

From all these results we can conclude that for harmonic $n=0$ the 
relaxation is non exponential at very short times and eventually becomes 
exponential for time larger than the Thouless time~:
\begin{eqnarray}
  \label{phasesp1}
  \smean{\EXP{\I\Phi}}_{V,{\cal C}_0}
  &\simeq&
  \exp{-\frac{\sqrt\pi}{4} \left(\frac{t}{\tau_N}\right)^{3/2} }
  \hspace{0cm}\mbox{ for } t\ll\tau_D\\
  \label{phasesp2}
  &\simeq&\exp{ -\frac{1}{6} \frac{t}{\tau_c} }
  \hspace{1.4cm}\mbox{ for } t\gg\tau_D
  \:.
\end{eqnarray}
This difference  comes from the time evolution of 
$\mean{W}_{{\cal C}_0}$~: when the ring has not been explored 
($t\ll\tau_D$) it scales like $\mean{W}_{{\cal C}_0}\sim\sqrt{t}$, 
while it becomes time independent for ergodic regime ($t\gg\tau_D$).

On the other hand the phase coherence relaxation is always exponential 
for harmonic $n\neq0$~:
\begin{eqnarray}
  \label{phasesp3}
  \smean{\EXP{\I\Phi}}_{V,{\cal C}_n}
  \simeq\exp{ -\frac{1}{6}\frac{t}{\tau_c} }
  \hspace{0.5cm} \forall\: t 
  \:.
\end{eqnarray}
This is due to the fact that the trajectories with finite winding necessarily
explore the ring which leads to $\mean{W}_{{\cal C}_n}\sim t^0$, for all
times, as explained in the introduction.

\mathversion{bold}
\subsection{Large perimeter $L\gg L_N$}
\mathversion{normal}

\subsubsection*{a. Short times}

In this regime the analysis provided for small perimeter using the results
of section \ref{sec:diffphase} remains valid. 
The condition of validity of the results slightly changes since the 
times are now in the following order~: $\tau_c\ll\tau_N\ll\tau_D$.
For harmonic $n=0$, eq.~(\ref{phasesp1}) holds for $t\ll\tau_N$.
For the harmonics $n\neq0$, eq.~(\ref{phasesp3}) holds for $t\ll\tau_c$.

\subsubsection*{b. Long times}

\noindent$\bullet$
\mathversion{bold}
{\bf Harmonic $n=0$.}
\mathversion{normal}
%
For this harmonic, the magnetoconductivity is given by AAK since for 
$L_N\ll L$ the boundary conditions are not important. The inverse Laplace 
transform of eq.~(\ref{AAKAiry}) has been computed exactly in
ref.~\cite{AkkMon04,MonAkk05,footnote4}~:
$
\smean{\EXP{\I\Phi}}_{V,{\cal C}_0} = 
\sqrt{\frac{\pi t}{\tau_N}}\sum_{n=1}^\infty \frac1{|u_n|}
\EXP{-|u_n|t/\tau_N}
$,
where $u_n$'s are zeros of ${\rm Ai}'(z)$.

\vspace{0.25cm}

\noindent$\bullet$
\mathversion{bold}
{\bf Harmonic $n\neq0$.}
\mathversion{normal}
%
In this case we can only use the method (C). 
We compare (\ref{harmgrandL}) to the integral (\ref{steepdes})~:
the dependence in $n$ of the exponential gives the exponent $\alpha=1$. 
Then its $L$-dependence gives $\Upsilon=\frac\pi8\frac{L}{L_N^3}$. 
The analysis of the prefactor shows that $\mu=0$.  Therefore~:
\begin{eqnarray}
  \smean{\EXP{\I\Phi}}_{V,{\cal C}_n}
  \simeq \exp{ -\frac{\pi^2}{64}\frac{t}{\tau_c} } 
  \:.
\end{eqnarray}

\subsubsection*{c. Summary}


For the harmonic $0$ \cite{AkkMon04,MonAkk05}~:
\begin{eqnarray}\label{GE1}
  \smean{\EXP{\I\Phi}}_{V,{\cal C}_0}
  &\simeq&\exp{-\frac{\sqrt\pi}{4}\left(\frac{t}{\tau_N}\right)^{3/2}} 
\mbox{ for } t\ll\tau_N \\
  \label{GE2}
  &&\hspace{-1.25cm} \simeq\sqrt{\frac{\pi t}{\tau_N}}\,\frac1{|u_1|}\, 
  \exp{-|u_1|\frac{t}{\tau_N}} 
  \hspace{0.25cm}
\mbox{ for } \tau_N\ll t
\end{eqnarray}
(the first zero of ${\rm Ai}'(z)$ is $|u_1|\simeq1.019$).

For harmonic $n\neq0$, we have seen above that~:
\begin{eqnarray}
  \smean{\EXP{\I\Phi}}_{V,{\cal C}_n}
  &\simeq&\exp{ -\frac{1}{6}\frac{t}{\tau_c} }
  \hspace{0.45cm}\mbox{ for } t\ll\tau_c\\
  &\simeq&\exp{ -\frac{\pi^2}{64}\frac{t}{\tau_c} } 
  \hspace{0.25cm}\mbox{ for } \tau_c\ll t 
  \:.
\end{eqnarray}

\mathversion{bold}
\subsection{From exponential phase coherence relaxation to 
            non exponential size dependent harmonics}
\mathversion{normal}

\begin{figure}[!ht]
\begin{center}
\includegraphics[scale=1]{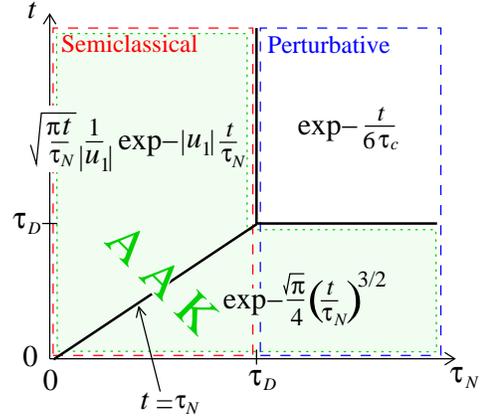}
\end{center}
\caption{(Color online) 
          {\it Relaxation of phase coherence for trajectories that do not
             wind around the flux~:} $\smean{\EXP{\I\Phi}}_{V,{\cal C}_0}$.
          {\it   The result for the infinite wire (AAK) is recovered when 
            trajectories cannot explore the whole ring, that is either
            for $t\ll\tau_D$ or for $t\ll\tau_N$.}
             \label{fig:dephasing1}}
\end{figure}

\begin{figure}[!ht]
\begin{center}
\includegraphics[scale=1]{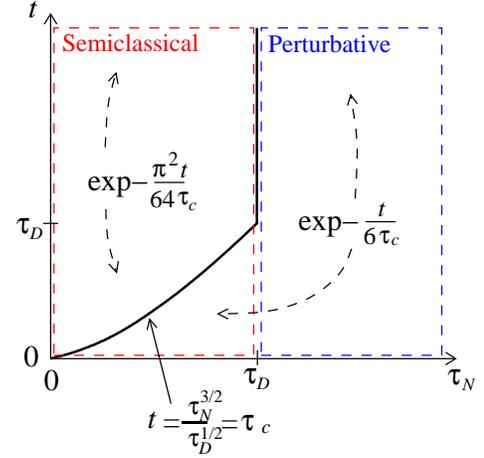}
\end{center}
\caption{(Color online)
          {\it Relaxation of phase coherence for trajectories with a finite
             winding number~:} $\smean{\EXP{\I\Phi}}_{V,{\cal C}_n}$, 
            {\it with}
             $n\neq0$. 
             \label{fig:dephasing2}}
\end{figure}

On figures~\ref{fig:dephasing1} and \ref{fig:dephasing2} we summarize the 
results obtained for the function $\smean{\EXP{\I\Phi}}_{V,{\cal C}_n}$.

The behaviour $\smean{\Phi^2}_{V}\propto{t}^{3/2}$ was first mentioned in
ref.~\cite{SteAhaImr90}, where it was conjectured that it may lead to
interesting effects in a ring. However, when the effect of winding is properly
taken into account, it turns out that the interesting effects in the ring come
from an exponential relaxation, {\it i.e.}
$\smean{\Phi^2}_{V,{\cal{C}}_n}\propto{t}$. In order to emphasize this point,
let us summarize the relationship between time dependence of the phase
relaxation and the decay of the harmonics. For $n\neq0$, the function
$\smean{\EXP{\I\Phi}}_{V,{\cal{C}}_n}$ is always exponential,
$\exp-\beta{t}/{\tau_c}$, with $\beta=1/6$ or $\beta={\pi^2}/{64}$, depending
on the time regime (see figure~\ref{fig:dephasing2}). The weak localization is
given by the time integrated probability to turn $n$ times around the ring
weighted by the exponential damping~:
\begin{eqnarray}
  \smean{\Delta\sigma_n} \sim \int_0^\infty\D t\,\EXP{-\beta t/\tau_c}\,
  \frac1{\sqrt t}\, \EXP{ -\frac{(nL)^2}{4t} }
  \:.
\end{eqnarray}
We recover the {\it non exponential} size dependence of
$\smean{\Delta\sigma_n}$, eqs.~(\ref{harmpetitL},\ref{harmgrandL}),
\begin{eqnarray}
  \label{behavLM}
  \hspace{-0.25cm}
  \smean{\Delta\sigma_n} 
  &\sim& 
  \exp{ -\sqrt\beta\,|n|\left(\frac{L}{L_N}\right)^{3/2} }
  \sim \EXP{-|n| L^{3/2}T^{1/2}}
  \!\!,
\end{eqnarray}
consequence of an {\it exponential} relaxation of phase coherence.

\section{The effect of connecting wires \label{sec:efconwires}}

Up to now we have considered an isolated ring. This was an 
important assumption in order to calculate  the path integral.
However, in a transport experiment the ring is necessarily connected to 
wires through which the current is injected. This has two important 
consequences that we now discuss.

\subsection{Classical nonlocality and quantum nonlocality}

\noindent$\bullet$ {\it Classical nonlocality.--}
The classical conductance of a wire of section $\sw$ and length
$L$ is given by the Ohm's law $G^{\rm cl}=\sigma_0\sw/L$ where $\sigma_0$ 
is the Drude conductivity. This result can be rewritten for
the dimensionless conductance as
$g^{\rm cl}_{\rm wire}=\frac{G^{\rm cl}}{2e^2/h}= \alpha_d N_c \ell_e/L$, 
where $N_c$ is
the number of channels, $\ell_e$ the elastic mean free path and
$\alpha_d$ a numerical constant depending on the dimension
($\alpha_1=2$, $\alpha_2=\pi/2$ and $\alpha_3=4/3$).
The quantum correction to the classical result is given by
\begin{eqnarray}\label{lefil} 
  \smean{\Delta g_{\rm wire}} = -\frac{2}{L^2} 
  \int_0^L \D x\, P_c(x,x) 
  \:,
\end{eqnarray}
where we have introduced the notation
$P_c(x,x')=\int_0^\infty\D t\,\EXP{-t/\tau_\varphi}\widetilde{\cal P}(x,x';t)$.

For a multiterminal network with arbitrary topology, the classical transport
is described by a conductance matrix that can be obtained by classical
Kirchhoff laws. This classical conductance matrix is a nonlocal object since
each matrix element depends on the whole network and the way it is connected
to external contacts. On such a network, because of the absence of translation
invariance, we have shown in ref.~\cite{TexMon04} how the cooperon must be
properly weighted when integrated over the wires of the network in order to
get the weak localization correction to the conductance matrix elements.
Eq.~(\ref{lefil}) generalizes as a sum of contributions of the different
wires~:
\begin{eqnarray}\label{RES3PRL} 
  \smean{\Delta g }= -\frac{2}{{\cal L}^2} \sum_{i}
  \frac{\partial{\cal L}}{\partial l_{i}}
  \int_{{\rm wire}\ i}\D x\, P_c(x,x) 
  \:,
\end{eqnarray} 
where ${\cal L}$ is the equivalent length obtained from Kirchhoff laws~:
$g^{\rm cl}=\alpha_d N_c \ell_e/{\cal L}$.  The weight of the wire $i$ is the
derivative of the equivalent length with respect to the length of the wire
$l_i$. The existence of these weights can lead to unexpected results, like a
change in sign of the weak localization correction for multiterminal
geometries~\cite{TexMon04,TexMon04b,TexMonAkk05}.

\begin{figure}[!ht]
\begin{center}
\includegraphics[scale=1]{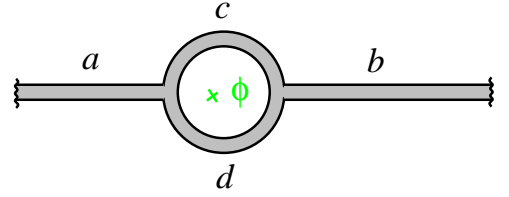}
\end{center}
\caption{ (Color online) 
         {\it A mesoscopic ring connected at two reservoirs (represented 
             by the wavy lines).\label{fig:loop2}}}
\end{figure}

When we consider the ring of figure~\ref{fig:loop2}, the equivalent length is
given by ${\cal L}=l_a+l_{c\parallel d}+l_b$ where
$l_{c\parallel{d}}^{-1}=l_c^{-1}+l_d^{-1}$. It follows that
\begin{eqnarray}\label{condweighted}
  &&\smean{\Delta g} = -\frac{2}{(l_a+l_{c\parallel d}+l_b)^2}
  \\\nonumber
&&\times
  \left[
    \int_a+\frac{l_d^2}{(l_c+l_d)^2}\int_c
    +\frac{l_c^2}{(l_c+l_d)^2}\int_d + \int_b
  \right]
  \D x \,P_c(x,x)
  \:.
\end{eqnarray}

\vspace{0.25cm}

\noindent$\bullet$ {\it Quantum nonlocality of the cooperon.--}
The cooperon is a nonlocal object that depends on the whole network.
$P_c(x,x)$ is a sum of contributions of diffusive loops that explore the
network over distances of order of the phase coherence length. We have shown
recently in refs.~\cite{TexMon04b,TexMon05} that the presence of the
connecting wires can strongly affect the behaviour of the harmonics of the AAS
oscillations.  We can distinguish two regimes for long connecting wires~:
({\it i}) in the limit $L_\varphi\ll L$ the AAS harmonics are exponential
$\smean{\Delta{g}_n}\propto\exp{-|n|L/L_\varphi}$. ({\it ii}) However in the
limit $L\ll L_\varphi$, the behaviour of the harmonics becomes
$\smean{\Delta{g}_n}\propto\exp{-|n|\sqrt{2L/L_\varphi}}$. This different
behaviour was analyzed in detail and shown to originate from the fact that the
Brownian trajectories can explore the connecting wires over distances larger
than the perimeter $L$.  The effective perimeter of a Brownian trajectory
encircling the ring is
$\ell_{\rm{eff}}\simeq\sqrt{2L/L_\varphi}\gg{L/L_\varphi}$.

\vspace{0.25cm}

We see that the simple exponential decay of the harmonics with the perimeter,
eq.~(\ref{AAS}), can be modified for two reasons~: either the presence of
connecting wires, or the effect of the electron-electron interaction. One acts
on the nature of the diffusion around the ring, the other acts on the nature
of the dephasing. In this section we propose to combine these
two effects.

The presence of connecting wires modifies the diffuson, and therefore the
function $W(x,x')$. The main difficulty to compute the path integral giving
the cooperon is that $W(x,x')$ can not be written as a single function of
$x-x'$ and it is not possible to make the path integral local in time.

\mathversion{bold}
\subsection{Large perimeter $L\gg L_N$}
\mathversion{normal}

The connecting wires are not expected to have a striking effect in this
regime. Since the cooperon vanishes exponentially over distances larger than
$L_N$ (or $L_\varphi$), the integration of the cooperon over the connecting
wires can be neglected when studying the harmonics. On the other hand the
diffuson is affected by the presence of connecting wires, and the function
(\ref{fctW}) has to be replaced by (\ref{Wr1},\ref{Wr3}). This function has a
similar structure to (\ref{fctW})~; moreover the two functions are equal in
the limit of long connecting wires as discussed in appendix~\ref{sec:gfr}. The
function was given in ref.~\cite{LudMir04} for a symmetric ring ($l_a=l_b$ and
$l_c=l_d$). In this case, when both $x$ and $x'$ are in the same arm of the
ring it reads~\cite{LudMir04}
$W(x,x')=\frac12|x-x'|(1-\frac{\gamma+1}{L}|x-x'|)$, where
$\gamma={l_{c\parallel d}}/{{\cal L}}$. This explains why, in the regime
$L\gg{L_N}$, the presence of connecting wires has almost no effect. It
essentially modifies the numerical prefactor in the exponential~:
$
\smean{\Delta g_n}\propto
\exp-n\frac{C_\gamma}{C_0}\frac\pi8(\frac{L}{L_N})^{3/2} 
$.  
The coefficient ${C_\gamma}/{C_0}$ interpolates smoothly between $1$ (for
$\gamma=0$) and $1/\sqrt2$ (for $\gamma=1$)~\cite{LudMir04}.

\vspace{0.25cm}

\noindent{\it Prefactor of the harmonics.--}
Before going to the limit of small perimeter we consider into more details the
prefactor of the conductance.  The exact result (\ref{RES4}) was derived for
an isolated ring and it is not clear how it is related to the conductance
through a ring connected to contacts by arms. In this latter case the
conductance is given by eq.~(\ref{condweighted}).
In the limit $L_N\ll l_a,l_b\ll l_c,l_d$, let us approximate the cooperon by
\begin{eqnarray} 
  \label{approx1}
  P_c(x,x) 
  &\simeq& P_c|_{\infty\ {\rm wire}} 
  \hspace{0.85cm} \mbox{ for } x\in a,b \\
  \label{approx2}
  &\simeq& P_c|_{{\rm isolated\ ring}} 
  \hspace{0.25cm} \mbox{ for } x\in c,d 
  \:,
\end{eqnarray}
where $P_c|_{\infty\ {\rm wire}}$ and $P_c|_{{\rm isolated\ ring}}$ are the
cooperon for an infinite wire and the isolated ring, respectively. This means
that we neglect the fact that the cooperon can leak in the wires $a$ and $b$,
when its coordinate is inside the ring. The harmonic $n=0$, given by
\begin{eqnarray}
  \label{sigapp1}
  \smean{\Delta g_0} 
  \simeq \frac{L_N}{{\cal L}}\,
  \frac{{\rm Ai}(L_N^2/L_\varphi^2)}{{\rm Ai}'(L_N^2/L_\varphi^2)}
  \:,
\end{eqnarray}
has the form $\smean{\Delta g_0}=(\pi/e^2)\smean{\Delta\sigma_0}/{\cal L}$.
The harmonics $n\neq0$ read
\begin{eqnarray}
  \label{sigapp2}
  \smean{\Delta g_n} 
  &\simeq& 
  \frac{l_{c\parallel d}\,L_N}{{\cal L}^2}\, 
  \frac{{\rm Ai}(L_N^2/L_\varphi^2)}{{\rm Ai}'(L_N^2/L_\varphi^2)}\:
  \EXP{-|n| \ell_{\rm eff} }
  \:,
\end{eqnarray}
where the effective perimeter is given by eq.~(\ref{lefflarpe}).

One may question the validity of the hypothesis (\ref{approx1},\ref{approx2}).
To answer this question, let us consider the limit 
$L_N/L_\varphi=\infty$ of 
eqs.~(\ref{sigapp1},\ref{sigapp2}) and compare
with the exact result obtained in ref.~\cite{TexMonAkk05}.
In eqs.~(\ref{sigapp1},\ref{sigapp2}), the ratio of Airy functions is 
replaced by $-L_\varphi/L_N$ and $\ell_{\rm eff}=L/L_\varphi$.
On the other hand the exact result gives
$\smean{\Delta g_0}\simeq-L_\varphi/{\cal L}$ and, 
for the harmonics $n\neq0$,
\begin{eqnarray}
  \label{eq:noLN1}
  \smean{\Delta g_n} 
  \simeq 
  -\frac{ l_{c\parallel d}\,L_\varphi}{ {\cal L}^2 }
  \left(\frac23\right)^{2|n|} \EXP{-|n|L/L_\varphi}
  \:.
\end{eqnarray}
The comparison between eq.~(\ref{eq:noLN1}) and eq.~(\ref{sigapp2}) for
$L_N/L_\varphi=\infty$ shows that we only missed the factor $(2/3)^{2|n|}$. 
This factor
is related to the probability for the diffusive trajectory to remain inside
the ring, when arriving at the vertex (see
refs.~\cite{AkkComDesMonTex00,TexMon05,TexMonAkk05}). Note that, in the limit
of short connecting wires $l_a\ll L_\varphi$, the factor $(2/3)^{2|n|}$ in
eq.~(\ref{eq:noLN1}) is replaced by
$[1+{L_\varphi^2}/({l_aL})]({2l_a}/{L_\varphi})^{|n|}$, which describes the
probability that the diffusive particule is not absorbed by the nearby
reservoir when arriving at the vertex. In the limit $l_a/L_\varphi\to0$, the
reservoirs break phase coherence at the vertices and the harmonics vanish.

\hspace{0.25cm}

Finally we remark that the prefactor of the harmonics differs from the 
one obtained by LM~\cite{LudMir04}. With our notations their result reads~:
$
\smean{\Delta g_n}_{\rm LM}
\sim L_N^{9/4}\,\EXP{-|n|(\frac{L}{L_N})^{3/2}}
$,
whereas our prefactor is linear in $L_N$ (for $L_\varphi=\infty$), as for
$\smean{\Delta\sigma_n}$. We stress that the difference between LM's and our
result is not due to the presence of connecting wires. The linear dependence
in $L_N$ of the prefactors of eqs.~(\ref{RES4},\ref{sigapp2}) comes from the
path integral. In LM's paper, the $L_N^{9/4}$ comes from a wrong
estimation of the prefactor of the path integral.
In appendix~\ref{sec:instanton} we have shown how the correct prefactor can be
extracted within the instanton approach followed by LM.

\mathversion{bold}
\subsection{Small perimeter $L\ll L_N$}
\mathversion{normal}

In this limit, the Brownian trajectories contributing to the path integral
(\ref{spAAK},\ref{spAAK2}) are related to times $t\gg\tau_D$. It is known that
for such time scales the arms have a striking effect since the diffusive
trajectories spend most of the time in the long connecting wires (see
ref.~\cite{TexMon05} and section 5.5 of ref.~\cite{ComDesTex05}).  This
affects both the winding properties around the ring and the nature of the
dephasing.


We expect that the relaxation of the phase coherence mainly occurs inside the
arms, {\it i.e.} the largest contribution to
$\smean{\Phi^2}_V\propto\int_0^t\D\tau\,W(x(\tau),x(\bar\tau))$ correspond to
$x$ and $x'$ in the arms. In this case the function $W(x,x')$ is given by
eqs.~(\ref{Wr4},\ref{Wr5}). If we consider the long arm limit
$l_a,l_b\gg{L_N}\gg{l_c},l_d$, we can take the limit $l_a,l_b\to\infty$ in
(\ref{Wr4},\ref{Wr5}) and we have~: $W(x,x')\simeq\frac12|x-x'|$. We recover
the same function as for the infinite wire.  In this limit the length over
which the trajectories extend in the wires is not limited by the size of the
system but by the time. Therefore we expect that the phase coherence
relaxation is non exponential and the function
$\smean{\EXP{\I\Phi}}_{V,{\cal{C}}_n}$ is similar to the one obtained for the
infinite wire (or for the harmonic $n=0$ for the large ring,
eqs.~(\ref{GE1},\ref{GE2}))~:
$
\smean{\EXP{\I\Phi}}_{V,{\cal{C}}_n}
\simeq\smean{\EXP{\I\Phi}}_{V,\mathcal{C}}|_{\rm wire}
=f_{\rm wire}(t/\tau_N)
$.
This funtion decreases monotoneously over the scale~$\tau_N$.




Since the cooperon $P_c(x,x)$ is expected to decay over a scale $L_N$ in the
arms, the integration over $x$ in eq.~(\ref{condweighted}) leads to 
(integration is dominated by the contributions of the arms)~:
\begin{equation}
  \label{eq100}
  \smean{\Delta g_n} \sim
  -\frac{L_N}{(l_a+l_b)^2}\int_0^\infty\D t\,\mathcal{P}_n(x,x;t)\,
  \smean{\EXP{\I\Phi}}_{V,\mathcal{C}_n}
\end{equation}
where $x$ is now a point inside the ring (in the regime
$t\gg\tau_D\propto{L}^2$, the probability 
$\mathcal{P}_n(x,x;t)$ is almost independent on $x$,
provided it remains at a distance smaller than $\sqrt{t}$ from the ring). The
probability to wind $n$ times around the connected ring is \cite{TexMon05}
$
\mathcal{P}_n(x,x;t)
\simeq\frac{\sqrt{L}}{2^{3/2}t^{3/4}}\,
\psi(\xi=\frac{n\sqrt{2L}}{t^{1/4}})
$
where $\psi(\xi)\simeq\frac4{\sqrt{6\pi}}(\xi/4)^{1/3}\EXP{-3(\xi/4)^{4/3}}$
for $\xi\gg1$ [$\psi(0)$ is finite~\cite{TexMon05}].
Therefore, as a function of time,
$\mathcal{P}_n(x,x;t)$ increases until $t\sim{n}^4\tau_D$ and then presents a
smooth tail $t^{-3/4}$. In order to evaluate eq.~(\ref{eq100}), we have now to
consider two limits~:

\noindent$\bullet$
For $n^2L\ll L_N$, the integral (\ref{eq100}) is dominated by the tail of the
probability $\mathcal{P}_n(x,x;t)$, therefore
\begin{equation}
  \smean{\Delta g_n} \sim
  -\frac{L_N^{3/2}L^{1/2}}{(l_a+l_b)^2}
\end{equation}

\noindent$\bullet$
For $n^2L\gg L_N$, only the tail of 
$f_{\rm wire}(t/\tau_N)\sim\sqrt{t/\tau_N}\exp-|u_1|t/\tau_N$ is important.
The steepest descent method gives~:
\begin{equation}
   \smean{\Delta g_n} \sim
  -\frac{L_N^{3/2}L^{1/2}}{(l_a+l_b)^2}
  \left(\frac{n^2L}{L_N}\right)^{7/12}\,
  \EXP{-\kappa|n|\sqrt{L/L_N}} 
\end{equation}
whith $\kappa=\sqrt2|u_1|^{1/4}\simeq1.421$. We recover a dependence of the
harmonics reminiscent to the one obtained for exponential relaxation of phase
coherence~\cite{TexMon05} (see discussion of paragraph~V.A). Finally, we
remark that the expected decay of the harmonics in temperature is~:
$\smean{\Delta g_n}\sim-T^{-11/36}\EXP{-nL^{1/2}T^{1/6}}$, for sufficiently
large~$n$.

This prediction should be tested experimentally on a chain of rings separated
by sufficiently long wires, compared to the phase coherence length (several
rings are required in order to perform a disorder average).


\begin{table}[!ht]
\begin{center}
\begin{tabular}{|c|c|}
\hline
   & {Exponential relaxation}      \\
\hline
$L\gg L_\varphi$ & $\exp-nL/L_\varphi$          \\[0.15cm]
$L\ll L_\varphi$ & $\exp-n(2L/L_\varphi)^{1/2}$ \\
\hline\hline
   & {Electron-electron interaction}    \\
\hline
$L\gg L_N$       & $\exp-\frac\pi8 n(L/L_N)^{3/2}$ \\ [0.15cm]
$L\ll L_N$       & $\exp-\kappa\,n(L/L_N)^{1/2}$\\
\hline
\end{tabular}
\end{center}
\caption{\it The harmonics $\smean{\Delta g_n}$ 
         of the conductance through a ring of perimeter $L$ connected 
         to long arms. We compare the results for exponential phase 
         coherence relaxation, described by $L_\varphi$, and the one obtained
         for electron-electron interaction ($L_N$).
         $\kappa$ is the dimensionless constant given in the text.}
\end{table}



\section{Conclusion}

We have considered the effect of the electron-electron interaction on the weak
localization correction for a diffusive ring.  We have calculated exactly the
path integral giving the weak localization correction for the isolated ring in
the presence of electron-electron interaction (characterized by the Nyquist
length $L_N$) and of other dephasing mechanisms described by an exponential
phase coherence relaxation (characterized by $L_\varphi$). The harmonics of
the conductivity are always of the form
$\smean{\Delta\sigma_n}\propto\exp-|n|\ell_{\rm eff}$, where $\ell_{\rm eff}$
accounts for both kinds of relaxation, combined in a nontrivial way.
The effective perimeter can always be written as~:
\begin{eqnarray}
  \ell_{\rm eff} = \left(\frac{L}{L_N}\right)^{3/2}
  f(L_c^2/L_\varphi^2)
  \:,
\end{eqnarray}
where $L_c=L_N^{3/2}/L^{1/2}$. For large perimeter $L\gg L_N$, the 
dimensionless function
is $f(x)=\eta(x)$ [eq.~(\ref{fctetaapp})]. For small perimeter 
$L\ll L_N$ it is given by
$f(x)=\sqrt{1/6+x}$. All limiting behaviours of $\ell_{\rm eff}$ have
been studied in sections \ref{sec:subsecSP} and~\ref{sec:subsecLP}.

In order to interpret these results, we have studied the function
$\smean{\EXP{\I\Phi}}_{V,{\cal C}_n}$ characterizing the phase coherence
relaxation for trajectories with winding $n$ (involved in the $n$-th harmonic
of the AAS oscillations). We have shown that, whereas the phase relaxation
crosses over from a non exponential behaviour to an exponential behaviour for
the harmonics $n=0$, it is always exponential for $n\neq0$. The time $\tau_c$
characterizing the exponential relaxation
$\smean{\EXP{\I\Phi}}_{V,{\cal{C}}_n}\sim\exp-t/\tau_c$ is given by~:
$\tau_c=\sigma_0\sw/(e^2TL)=\tau_N^{3/2}\tau_D^{-1/2}$.  This exponential
relaxation is at the origin of the non exponential decay of the harmonics with
the size and the new temperature dependence~\cite{footnote5} predicted by LM~:
$\smean{\Delta{g}_n}\sim\exp-n(L/L_N)^{3/2}\sim\exp-nT^{1/2}L^{3/2}$.

In the light of these results it seems  difficult to interpret
the experiment recently performed on large GaAs/GaAlAs square networks where a
behaviour $\exp{-nT^{1/3}}$ was observed
\cite{FerAngRowGueBouTexMonMai04,Fer04}. However the experiment was performed
in a temperature range where $L_N\sim{L}$. In this regime the diffusive
trajectories start to explore the network surrounding the loop, which modifies
the behaviour of the harmonics as a function of $L/L_N$ and therefore the
temperature dependence.
However we have not been able to extend the theory to the square network. 

In the last part of our article we have studied the effect of the wires
connecting the ring to reservoirs. Whereas the AAS harmonics are 
weakly affected by the connecting wires in the limit $L\gg{L_N}$, we 
have shown that a strong modification is expected in the opposite limit
$L\ll{L_N}$, where we have predicted a behaviour
$\smean{\Delta{g}_n}\sim{T}^{-11/36}\exp-nL^{1/2}T^{1/6}$.
The appropriate experimental setup to test this result is a chain of rings
separated by wires whose length remain larger than $L_N$.
This situation is particularly interesting because the flux sensitivity
is due to the motion inside the ring whereas, in this case, the dephasing
occurs mostly in the arms.

An interesting effect has been observed recently in the study of four terminal
measurements of AB oscillations in a ballistic ring. It has been shown
experimentally \cite{KobAikKatIye02} that the dephasing rate depends on the
configuration of the voltage probes and current probes. 
It was suggested that a measurement with current probes on both sides of the
ring favors charge fluctuations inside the ring and leads to a high dephasing
rate, whereas a nonlocal measurement with current probes at one side and
voltage probes at the other side ({\it i.e.} no current flows through the ring
on average) diminishes charge fluctuations in the ring and therefore leads to
smaller dephasing rate. This effect has been described theoretically in
ref.~\cite{SeePilJorBut03}. An interesting question is whether a similar
effect might occur in a diffusive ring.


\section*{Acknowledgments}

It is our pleasure to acknowledge Christopher B\"auerle, Eug\`ene Bogomolny,
H\'el\`ene Bouchiat, Markus B\"uttiker, Meydi Ferrier, Sophie Gu\'eron,
Alistair Rowe, Laurent Saminadayar and F\'elicien Schopfer for stimulating
discussions.


\begin{appendix}

\mathversion{bold}
\section{The function $W(x,x')$\label{sec:gfr}}
\mathversion{normal}

\subsection{Isolated ring}

We give here the solution of the equation 
$(\gamma-\Dc_x^2)P(x,x')=\delta(x-x')$ on a ring pierced by a flux.
$\Dc_x=\D/\D x-\I\theta/L$ is the covariant derivative. We introduce the 
variable $\chi=x/L\in[0,1]$. The solution of the 
equation $(b-\Dc_\chi^2)C(\chi,\chi')=\delta(\chi-\chi')$ with
$b=\gamma L^2$ and $\Dc_\chi=\D/\D\chi-\I\theta$ is given by 
(\ref{expressionC}) for 
$f(\chi)=\frac{\sinh\sqrt{b}(1-\chi)}{\sinh\sqrt{b}}$. 
The two Green's functions are related by  $P(x,x')=L\,C(x/L,x'/L)$.

\vspace{0.25cm}

\noindent $\bullet$ $\gamma=0$ \& $\theta\neq0$.--
Let us consider the limit $\gamma=0$. We have~: 
\begin{eqnarray}
  C(\chi,\chi') &=& \frac{\EXP{\I\theta(\chi-\chi')}}{2(1-\cos\theta)}
\\\nonumber
 &&\hspace{-0.5cm}\times
  \left[ 
    1 + |\chi-\chi'|\left(\EXP{-\I\theta\sign(\chi-\chi')}-1\right)
  \right]
  \:.
\end{eqnarray}

\vspace{0.25cm}

\noindent $\bullet$ $\gamma=0$ \& $\theta=0$.--
In the limit of vanishing $\gamma$ and $\theta$, the diffusion equation
possesses a zero mode, therefore the Green's function $C(\chi,\chi')$ presents
a diverging contribution~:
$
C(\chi,\chi')
{ {\raisebox{-.3cm}{$\textstyle=$}} \atop {\scriptstyle{\theta\to0}} }
\frac1{\theta^2} - \frac12|\chi-\chi'|\left(1-|\chi-\chi'|\right)
.$
This diverging contribution disappears when considering the function 
\begin{eqnarray}
  \label{eqA2}
  W(x,x')&=&\frac{P(x,x)+P(x',x')}{2} -P(x,x') \\
  \label{Wr0}
  &=& \frac12|x-x'|\left(1-\frac{|x-x'|}{L}\right)
  \:.
\end{eqnarray}

The existence of a zero mode is an artefact coming from the fact that the
system is isolated. In a more realistic situation (when the ring is connected
to reservoirs through wires, for example) the Laplace operator does not
possess a zero mode~\cite{AkkComDesMonTex00}. Physically, the zero mode does
not contribute to the function $W(x,x')$, {\it i.e.} to the dephasing, since
it corresponds to uniform fluctuations of the electric potential that do not
contribute to the phase~$\Phi$, given by eq.~(\ref{phase}).

\subsection{Connected ring}

\begin{figure}[!ht]
  \centering
  \includegraphics[scale=1]{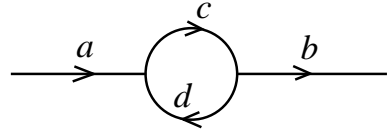}
  \caption{Orientation of the arcs of the ring.}
  \label{fig:loop6}
\end{figure}

We now construct the symmetric function $W(x,x')$, defined in
eqs.~(\ref{defW},\ref{eqA2}), when the ring is connected to reservoirs by two
wires (figure~\ref{fig:loop2}). In this case $W(x,x')$ is fully characterized
by a set of components corresponding to the coordinates in the different
wires. A system of coordinate must be specified~: we first give an orientation
to the wires of the network, shown on figure~\ref{fig:loop6} (we call ``arc''
an oriented wire). Then the coordinate along an arc $i$ belongs to the
interval $[0,l_i]$, where $l_i$ is the length of the arc. Below we construct
$W(x,x')$ when $x$ and $x'$ are both in the ring or both in the arms.

\subsubsection{Inside the ring}

When both coordinates are in the arc $c$, we obtain~:
\begin{eqnarray}
  \label{Wr1}
  W_{c,c}(x,x') = \frac12|x-x'|
  \left(
    1-\frac{l_a+l_d+l_b}{l_a+l_{c\parallel d}+l_b}\frac{|x-x'|}{l_c+l_d}
  \right)
  \:,
\end{eqnarray}
with $x,x'\in[0,l_c]$.
For $l_c=l_d$ we recover the expression given in ref.~\cite{LudMir04}.
We also need the case when $x\in{c}$ and $x'\in{d}$~:
\begin{eqnarray}
  \label{Wr2}
  W_{c,d}(x,x') &=& \frac12
  \frac{(l_a+l_b)l_{c\parallel d}}{l_a+l_{c\parallel d}+l_b}
  \left(
    1-\frac{x}{l_c}-\frac{x'}{l_d}
  \right)^2
  \nonumber\\&&\hspace{-0.25cm}
  +\frac12\left[
    x\left(1-\frac{x}{l_c}\right)
    +x'\left(1-\frac{x'}{l_d}\right)
  \right]
  \:,
\end{eqnarray}
with $x\in[0,l_c]$, $x'\in[0,l_d]$ and with the orientation of
figure~\ref{fig:loop6}. It is more convenient to consider a unique way to
measure both coordinates. Therefore we shift $x'$ by $l_c$ in~(\ref{Wr2})~:
\begin{widetext}
\begin{eqnarray}
  \label{Wr3}
  W_{c,d}(x,x'-l_c) &=& \frac12
  \bigg[
    -\frac{l_{c}^2}{l_a+l_{c\parallel d}+l_b}
    -\frac{l_a-l_{c\parallel d}+l_b}{l_a+l_{c\parallel d}+l_b} x
    +\left(1+\frac{2\,l_c^2}{(l_c+l_d)(l_a+l_{c\parallel d}+l_b)}\right)x'
    \nonumber\\&&
    -\frac{l_a+l_d+l_b}{l_a+l_{c\parallel d}+l_b} \frac{x^2}{l_c+l_d}
    -\frac{l_a+l_c+l_b}{l_a+l_{c\parallel d}+l_b} \frac{x'^2}{l_c+l_d}
    +\frac{l_a+l_b}{l_a+l_{c\parallel d}+l_b} \frac{2xx'}{l_c+l_d}
  \bigg]
  \:,
\end{eqnarray}
\end{widetext}
with $x\in[0,l_c]$ and $x'\in[l_c,l_c+l_d]$. It is now clear that in the limit
of long connecting wires, $l_a,l_b\gg l_c,l_d$, eqs.~(\ref{Wr1},\ref{Wr3})
lead to the same result as in the isolated ring, eq.~(\ref{Wr0}). We see that,
inside the ring, $W(x,x')$ is not everywhere a function of $x-x'$ only, apart
in the limit $l_a,l_b\to\infty$.

\subsubsection{Inside the arms}

When both coordinates are in the same arm, we have~:
\begin{eqnarray}
  \label{Wr4}
  W_{a,a}(x,x') = \frac12|x-x'|
  \left(
    1-\frac{|x-x'|}{l_a+l_{c\parallel d}+l_b}
  \right)
\end{eqnarray}
When the two coordinates belong to different arcs we prefer to shift the 
origin of the coordinate $x'$, as we did inside the ring. If the shift
is chosen to be $l_a+l_{c\parallel d}$, we obtain the simple expression~:
\begin{eqnarray}
  \label{Wr5}
  W_{a,b}(x,x'-l_a-l_{c\parallel d}) = \frac12(x'-x)
  \left(
    1-\frac{x'-x}{l_a+l_{c\parallel d}+l_b}
  \right)
\end{eqnarray}
(in this expression $x'=l_a+l_{c\parallel d}$ corresponds to the begining of
the arc $b$).  When $l_{c\parallel{d}}=0$ (no ring) we obtain from
(\ref{Wr4},\ref{Wr5}) the result for a connected wire
$W(x,x')=\frac12|x-x'|(1-\frac{|x-x'|}{l_a+l_b})$, which is similar to the one
for the isolated ring eq.~(\ref{Wr0}), as mentioned above. It is remarkable
that, in the presence of the ring, there exists a choice of coordinates for
which $W(x,x')$ in the arms has precisely the same structure as in the absence
of the ring. In the limit of an infinite wire, $l_a,l_b\to\infty$, we recover
from (\ref{Wr4},\ref{Wr5}) the result of the infinite wire,
$W(x,x')\simeq\frac12|x-x'|$.


\section{Hermite function\label{sec:Hermite}}

Consider the Hermite equation \cite{NikOuv83}~:
\begin{eqnarray}
  y''(z) -2z\,y'(z) + 2\nu\,y(z)=0
  \:.
\end{eqnarray}
Two linearly independent solutions are the Hermite function $H_\nu(z)$ 
and $H_\nu(-z)$.
An integral representation is
\begin{eqnarray}
  \label{intrep0}
  H_\nu(z) =\frac1{\Gamma(-\nu)} \int_0^\infty\frac{\D t}{t^{\nu+1}}\,
  \EXP{-t^2-2zt}
  \:,
\end{eqnarray}
from which we get the series representation~:
\begin{eqnarray}
  \label{seriesrep}
  H_\nu(z) = \frac1{2\Gamma(-\nu)}\sum_{n=0}^{\infty}
  (-1)^n\Gamma\left(\frac{n-\nu}{2}\right)\frac{(2z)^n}{n!}
  \:.
\end{eqnarray}

\vspace{0.25cm}

We now study several limiting behaviours of the Hermite function 
$H_{-\frac12+\I\omega}(\EXP{\I\pi/4}a^{1/4}(\chi-1/2))$, where 
$\omega=\frac12(\frac{b}{\sqrt{a}}+\frac{\sqrt{a}}{4})$.

\mathversion{bold}
\subsection{The limit $a\to0$}
\mathversion{normal}

In this case $\omega\simeq\frac{b}{2\sqrt{a}}$, therefore we study
the limit $\omega\to\infty$ when the argument of the Hermite function
reads $z=x/\sqrt{-\I\omega}$ with $x$ finite. Using the expression 
$
\Gamma\left(\frac{n}{2}+\frac14-\I\frac{\omega}{2}\right)
\propto\left(\frac{\omega}{2}\right)^{\frac{n}{2}}\EXP{-\I n\frac\pi4}
$,
valid for $\omega\gg n$, and the series representation (\ref{seriesrep}), 
we get
\begin{eqnarray}\label{asy1}
  H_{-\frac12+\I\omega}\left(\frac{x}{\sqrt{-\I\omega}}\right)
  \PROPTO{\omega\to\infty}
  \EXP{-\sqrt2\,x}
  \:.
\end{eqnarray}

\mathversion{bold}
\subsection{The limit $a\to\infty$~: from Hermite to Airy function}
\mathversion{normal}

The first step to study this limit is to perform a rotation of $+\pi/4$ 
in the complex plane of the axis of integration in (\ref{intrep0}).
One obtains
\begin{eqnarray}\label{intrep}
  \hspace{-0.25cm}
  H_{-\frac12+\I\omega}\left(\EXP{\I\pi/4}A\right)
  =\frac{\EXP{\I\frac\pi8+\frac\pi4 \omega}}
          {\Gamma\left(\frac12-\I\omega\right)}
  \int_0^\infty\frac{\D x}{\sqrt x}\,\EXP{-\I\varphi(x,\chi)}
\end{eqnarray}
where the phase reads
\begin{eqnarray}
  \varphi(x,\chi) = x^2 + 2A\,x+\omega\ln x
  \:.
\end{eqnarray}
We introduced the notation  
\begin{eqnarray}
  A=a^{1/4}\left(\chi-\frac12\right)
  \:.
\end{eqnarray}
We are interested in the limit $a\to\infty$ with $L_N/L_\varphi$ finite 
(or zero), 
therefore it is convenient to write~:
\begin{equation}
  \omega = \frac{\sqrt{a}}{8} \left( 1+4\frac{\Lambda}{a^{1/3}} \right)  
\end{equation}
where $\Lambda = ({L_N}/{L_\varphi})^2$.

\mathversion{bold}
\subsubsection*{({\it i}) The case $\chi\to0$}
\mathversion{normal}

The function $\varphi(x,\chi)$ is a monotonous function of the variable
$x\in\RR^+$, however its second derivative vanishes at $x=\pm\sqrt{\omega/2}$.
For $\chi\to0$ the first derivative at this point becomes very small in the
limit $a\to\infty$, therefore we expect that the neighbourhood of
$\sqrt{\omega/2}$ brings the dominant contribution to the integral. The
expansion of the phase in the neighbourhood of $\sqrt{\omega/2}$ reads (for
$t\to0$)
\begin{eqnarray}
  \varphi\left(\sqrt{\omega/2}+a^{1/12}t,\chi\right)
  &=&\varphi\left(\sqrt{\omega/2},\chi\right) 
\\\nonumber
  &&\hspace{-4cm}+\left(a^{1/3}\chi+\Lambda\right)t 
  +\frac13 t^3 - \frac{1}{2a^{1/6}}t^4+\cdots
  +O(a^{\frac{3-n}{6}}t^n)
  \:.
\end{eqnarray}
Inserting this expression into the integral representation (\ref{intrep}),
we obtain the Airy function \cite{AbrSte64}
${\rm Ai}(x) = \frac1{2\pi}\int_{-\infty}^{+\infty}\D t\,
\EXP{\I(\frac{t^3}3+xt)}$, namely~:
\begin{eqnarray}\label{fH2Ai0}
  &&  H_{-\frac12+\I\omega}\left(\EXP{\I\pi/4}a^{1/4}(\chi-\frac12)\right)
\nonumber\\ &&\hspace{1cm}  
  \PROPTO{a\to\infty}
  \EXP{-\I\frac{\sqrt{a}}{2}\chi}\: 
  {\rm Ai}\left(\Lambda+a^{1/3}\chi\right) 
  \:.
\end{eqnarray}
This expression is valid for $\chi\to0$ such that 
$a^{1/3}\chi$ is not large, and for $\Lambda/a^{1/3}\ll1$. This last condition
rewrites $L_\varphi\gg L_N\sqrt{L_N/L}$.

\mathversion{bold}
\subsubsection*{({\it ii}) The case $\chi\to1$}
\mathversion{normal}

\begin{figure}[!ht]
\begin{center}
\includegraphics[scale=0.8]{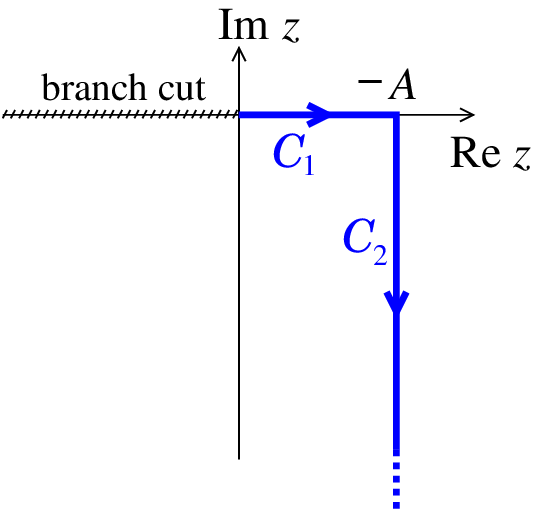}
\hspace{0.25cm}
\includegraphics[scale=0.8]{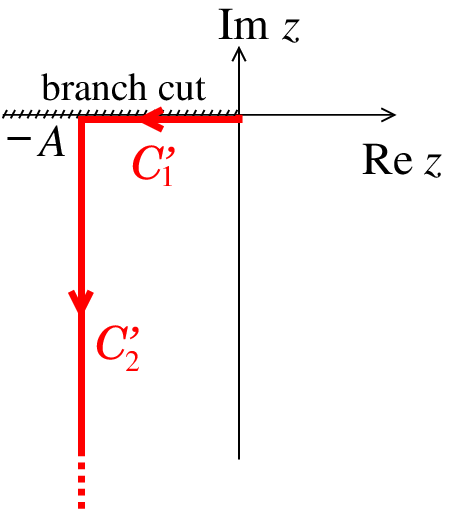}
\end{center}
\caption{(Color online)
         Left~:
         {\it Contour of integration in the complex plane for $\chi<1/2$ 
         ({\it i.e.} $A<0$).
         The vertical line separates the regions where $\re[\I(z^2+2Az)]$
         is positive or negative in the lower half of the complex plane.}
         Right~:
         {\it Contour of integration for $\chi>1/2$ 
         ({\it i.e.} $A>0$)}
         \label{fig:contour1}}
\end{figure}

For $\chi\to1$ the expansion of $\varphi(z,\chi)$ must be realized in the
neighbourhood of $-\sqrt{\omega/2}$, where the first derivative with respect
to $z$ vanishes (the first derivative at $+\sqrt{\omega/2}$ now diverges in
the limit $a\to\infty$). Therefore the contour of integration must be deformed
in order to visit the neighbourhood of $z=-\sqrt{\omega/2}$.  The new contour
of integration is shown on the right part of figure~\ref{fig:contour1}. For
$\chi>1/2$~:
\begin{eqnarray}
  \int_0^\infty\frac{\D x}{\sqrt x}\,\EXP{-\I\varphi(x,\chi)}
  =\int_{{\cal C}'_1+{\cal C}'_2}\frac{\D z}{\sqrt z}\,\EXP{-\I\varphi(z,\chi)}
  \:.
\end{eqnarray}
To deal with more symmetric expressions for $\chi>1/2$ and $\chi<1/2$ we
remark that the contour of integration can also be deformed in this latter
case (see the left part of figure~\ref{fig:contour1}). For $\chi<1/2$~:
\begin{eqnarray}
  \int_0^\infty\frac{\D x}{\sqrt x}\,\EXP{-\I\varphi(x,\chi)}
  =\int_{{\cal C}_1+{\cal C}_2}\frac{\D z}{\sqrt z}\,\EXP{-\I\varphi(z,\chi)}
  \:.
\end{eqnarray}

The dominant contribution to the integral is given by the 
contribution of the segment ${\cal C}'_1$.
By noting that
$\varphi(z\EXP{-\I\pi},\chi) = -\I\pi\omega+\varphi(z,1-\chi)$
for $\chi>1/2$ and $z\in\RR^+$, we see that, for $\chi>1/2$,
\begin{eqnarray}
  \int_{{\cal C}'_1}\frac{\D z}{\sqrt z}\,\EXP{-\I\varphi(z,\chi)}
  =-\I\EXP{-\pi\omega}
  \int_{{\cal C}_1}\frac{\D z}{\sqrt z}\,\EXP{-\I\varphi(z,1-\chi)}
  \:.
\end{eqnarray}
Therefore, since $\tilde f(\chi)$ is dominated by ${\cal C}_1$ for 
$\chi\to0$ and by ${\cal C}_1'$ for 
$\chi\to1$ we have
\begin{eqnarray}\label{0rel1}
  \tilde f(\chi) \simeq -\I\EXP{-\pi\omega}\tilde f(1-\chi)
  \hspace{0.5cm}\mbox{for } \chi\to1
  \:.
\end{eqnarray}


\section{Semiclassical approach\label{sec:instanton}}

In this appendix we analyze the cooperon solution of (\ref{eqcooperon}) by 
following a semiclassical approach, valid for $L_N\ll L$.
We construct the solution of the equation 
$[-\frac{\D^2}{\D\chi^2}+V(\chi)]f(\chi)=0$ for $\chi\in[0,1]$, 
where the potential is 
\begin{eqnarray}
  V(\chi)=a\chi(1-\chi)+b
  \:.
\end{eqnarray} 
The solution of interest satisfies $f(0)=1$ and $f(1)=0$.
As shown in the text, the harmonics of the cooperon then read
\begin{eqnarray}
C^{(n)}(0,0)=\frac{1}{2\sqrt{f'(0)^2-f'(1)^2}}\,\EXP{-|n|\ell_{\rm eff}}
\end{eqnarray}
with $\cosh\ell_{\rm eff}=\frac{f'(0)}{f'(1)}$.

The semiclassical approach holds for $a\gg1$. In this case the quadratic 
potential acts as a high barrier in which the solution hardly penetrates. 

\vspace{0.25cm}

\noindent{\bf Semiclassical solution~: instanton.--}
The WKB solution of the differential equation can be expressed as 
$f(\chi)\simeq\frac1{\sqrt{\pi(\chi)}}\exp\pm\int^\chi\D\chi'\,\pi(\chi')$
where the conjugate momentum for zero ``energy'' is 
$\pi(\chi)=\sqrt{V(\chi)}$.
Let us introduce~:
\begin{eqnarray}
  {\cal S}(\chi) = \int_0^\chi\D\chi'\,\sqrt{V(\chi')}
  \:,
\end{eqnarray}
which is the action of the instanton penetrating inside the potential 
barrier (classical solution for imaginary time) for zero ``energy''.
We write the semiclassical solution in the form
\begin{eqnarray}\label{f2}
  f(\chi) = \frac{a^{1/6}}{\sqrt{\pi}\,[V(\chi)]^{1/4}}\,
  \left( B_{\rm sc}\, \EXP{-{\cal S}(\chi)} 
       + C_{\rm sc}\, \EXP{+{\cal S}(\chi)}\right)
  \:,
\end{eqnarray}
where $B_{\rm sc}$ and $C_{\rm sc}$ are two coefficients to be determined
in order to satisfy boundary conditions.

\vspace{0.25cm}

\noindent{\bf Validity of the semiclassical approximation.--}
The validity of the semiclassical approximation can be expressed
precisely by the condition
$|\frac{\D}{\D\chi} \frac1{\pi(\chi)}| \ll1$.
This condition rewrites here as
\begin{eqnarray}\label{validity}
  a^{1/3} \left( \chi(1-\chi)+\frac{b}{a} \right) \gg1
  \:.
\end{eqnarray}
Depending on the relative magnitude of $L_N$ and $L_\varphi$, there are 
two possibilities that we discuss now.
We recall the notation $\Lambda=b/a^{2/3}=L_N^2/L_\varphi^2$

\mathversion{bold}
\subsection{Regime $L_\varphi\ll L_N$}
\mathversion{normal}

In this case $b/a^{2/3}\gg1$ and the condition (\ref{validity}) is fulfilled
for any $\chi\in[0,1]$.
We immediately find the coefficients $B_{\rm sc}$ and $C_{\rm sc}$ by 
imposing the boundary conditions and obtain~:
\begin{eqnarray}
  f(\chi) = \sqrt{ \frac{\pi(0)}{\pi(\chi)} } \:
  \frac{ \sinh[{\cal S}(1)-{\cal S}(\chi)] }{ \sinh{\cal S}(1) }  
  \:.
\end{eqnarray}
Therefore
\begin{eqnarray}
  \label{fp0re1} f'(0) &=& -\frac{a}{4b} -\sqrt{b}\, \coth {\cal S}_{\rm inst}
  \simeq -\sqrt{b} \\
  \label{fp1re1} f'(1) &=& -\frac{\sqrt{b}}{\sinh{\cal S}_{\rm inst}}
  \simeq -2\sqrt{b}\,\EXP{-{\cal S}_{\rm inst}}
  \:,
\end{eqnarray}
where we have introduced the notation ${\cal S}_{\rm inst}\equiv{\cal S}(1)$.
The effective perimeter simply reads~:
\begin{eqnarray}
  \ell_{\rm eff} = \argcosh\frac{f'(0)}{f'(1)} \simeq {\cal S}_{\rm inst}
  \:.
\end{eqnarray}

\mathversion{bold}
\subsection{Arbitrary $L_\varphi$}
\mathversion{normal}

In the case $L_\varphi\gtrsim L_N$ ({\it i.e.}  $b/a^{2/3}\lesssim1$) the
condition (\ref{validity}) cannot be fulfilled near the edges of the interval.
For $\chi\sim0$ the condition (\ref{validity}) can only be satisfied for
$\chi\gtrsim\chi_c$ whereas for $\chi\sim1$ it is satisfied for
$1-\chi\gtrsim\chi_c$.  We have defined $\chi_c$ by the two conditions
$a^{1/3}\chi_c+\Lambda\gg1$ and $\chi_c\ll1$ (the breakdown of the
semiclassical approximation near the edges can be simply understood by noting
that, for $b=0$, $\chi=0$ and $\chi=1$ are the turning points of the classical
solution for imaginary time for a zero ``energy''). Therefore the resolution
of the differential equation must be performed carefully near the edges.
We separate the interval $[0,1]$ into three parts~:

\vspace{0.25cm}

\noindent
({\it i}) In the neighbourhood of $0$ where the potential is linear,
the solution is a combination of two Airy functions~\cite{AbrSte64}~:
\begin{eqnarray}\label{f1}
  f(\chi) = 
  B_1\,{\rm Ai}(a^{1/3}\chi+\Lambda) +
  C_1\,{\rm Bi}(a^{1/3}\chi+\Lambda)
  \:.
\end{eqnarray}

\vspace{0.25cm}

\noindent
({\it ii}) {\it Semiclassical solution.--} Sufficiently far from the edges of
the interval, that is in the interval $[\chi_c,1-\chi_c]$, we can use the WKB
solution (\ref{f2}).

\vspace{0.25cm}

\noindent
({\it iii}) In the neighbourhood of $\chi=1$ the potential is almost linear 
again and the solution reads
\begin{eqnarray}\label{f3}
  f(\chi) &=& 
  B_2\,{\rm Ai}(a^{1/3}(1-\chi)+\Lambda) 
\nonumber\\
&&  
  + C_2\,{\rm Bi}(a^{1/3}(1-\chi)+\Lambda)
  \:.
\end{eqnarray}

\vspace{0.25cm}

The solution $f(\chi)$ is continuous and differentiable. Therefore the 
matching of the three expressions should now be performed in the 
region $\chi\sim\chi_c$ and $\chi\sim1-\chi_c$. It is clear from the 
definition of $\chi_c$ that the matching is realized in the  Airy functions'
asymptotic region. We obtain the  following relations between the 
coefficients~:
\begin{eqnarray}
  \left(\begin{array}{c}B_2 \\ C_2\end{array}\right) = 
  \left(\begin{array}{cc}0 & q \\ 1/q & 0\end{array}\right) 
  \left(\begin{array}{c}B_1 \\ C_1\end{array}\right) 
  \:,
\end{eqnarray}
where 
\begin{eqnarray}
  q=2\,\EXP{{\cal S}_{\rm inst}+\frac43\Lambda^{3/2}}\,\gg1
  \:.
\end{eqnarray}
We add to these relations the conditions
\begin{eqnarray}
  f(0) &=& B_1\,{\rm Ai}(\Lambda)+ C_1\,{\rm Bi}(\Lambda) = 1 \\
  f(1) &=& B_2\,{\rm Ai}(\Lambda)+ C_2\,{\rm Bi}(\Lambda) = 0
  \:.
\end{eqnarray}
Solving these equations we find~:
\begin{eqnarray}
  B_1 &=& \frac{ q^2\,{\rm Ai} }{ q^2\,{\rm Ai}^2- {\rm Bi}^2 } \\
  C_1 &=& \frac{ -{\rm Bi} }{ q^2\,{\rm Ai}^2- {\rm Bi}^2}
\end{eqnarray}
where Airy functions are taken at $\Lambda$.
We eventually find~:
\begin{eqnarray}
  f'(0) &=& a^{1/3}\, 
  \frac{ q^2\,{\rm Ai}\,{\rm Ai}' - {\rm Bi}\,{\rm Bi}' }
       { q^2\,{\rm Ai}^2- {\rm Bi}^2 }
  \simeq a^{1/3}\, \frac{ {\rm Ai}' }{ {\rm Ai} } \\
  f'(1) &=& -a^{1/3}\, \frac{ q/\pi }{ q^2\,{\rm Ai}^2- {\rm Bi}^2 }
  \simeq -a^{1/3}\frac{1}{\pi\, q\,{\rm Ai}^2}
  \:, 
\end{eqnarray}
where we have used that the Wronskian of the Airy functions is
${\cal W}[{\rm Ai},{\rm Bi}]={\rm Ai}\,{\rm Bi}'-{\rm Ai}'\,{\rm Bi}=1/\pi$
(see ref.~\cite{AbrSte64}). 
Finally the two derivatives are
\begin{eqnarray}\label{fp01re2}
  f'(0) &\simeq& a^{1/3}\, \frac{{\rm Ai}'(\Lambda)}{ {\rm Ai}(\Lambda) }
  \\\label{fp01re2b}
  f'(1) &\simeq& -\frac{a^{1/3}}{2\pi}\,
  \frac{\EXP{-{\cal S}_{\rm inst}-\frac43\Lambda^{3/2}}}{{\rm Ai}(\Lambda)^2} 
  \:.
\end{eqnarray}
We can check that (\ref{fp01re2},\ref{fp01re2b}) coincide with
(\ref{fp0re1},\ref{fp1re1}) in the limit $L_\varphi\ll L_N$. The effective
perimeter is given by
\begin{eqnarray}\label{leffinstant}
  \ell_{\rm eff} = {\cal S}_{\rm inst}
  + \ln
  \left(
    -4\pi {\rm Ai}(\Lambda){\rm Ai}'(\Lambda)\EXP{\frac43\Lambda^{3/2}}
  \right)
  \:.
\end{eqnarray}
Eq.~(\ref{fp01re2}) corresponds to the limit (\ref{eq55}) derived directly
from the exact solution, which gives the prefactor of the harmonics
(\ref{RES4}).

\vspace{0.25cm}

\noindent{\bf Action of the instanton.--}
We analyze more into detail the action corresponding to the crossing of the 
barrier~:
\begin{eqnarray}
  {\cal S}(1) &=& \int_0^1\D\chi\,\sqrt{V(\chi)} \\
  &&\hspace{-1cm}= 
  \pi\omega \left(1 - \frac2\pi \arcsin\frac1{\sqrt{1+{a}/({4b})}} \right) 
  +\frac{\sqrt{b}}{2}
\end{eqnarray}
where we recall that $\omega=\frac{\sqrt{a}}{8}+\frac{b}{2\sqrt{a}}$.
The action can be written in the form
\begin{eqnarray}
  \label{eq:C25}
  {\cal S}_{\rm inst} \equiv {\cal S}(1) = \sqrt{a}\ \eta({b}/{a})
\end{eqnarray}
where the function $\eta(x)$ is given by~:
\begin{eqnarray}
  \label{fctetaapp}
  \eta(x) = 
  \left(\frac{1}{4}+{x}\right)\arctan\frac1{\sqrt{4x}} 
  +\frac{\sqrt{x}}{2}
  \:.
\end{eqnarray}
This function presents the following limiting behaviours~:
\begin{eqnarray}
  &&\hspace{-1.25cm}
  \eta(x) =
  \pi\left(\frac{1}{8}+\frac{x}{2}\right)
  -\frac43\, x^{3/2} + O(x^{5/2}) 
  \mbox{ for }x\ll1
\\
  &&\hspace{-1.25cm}
  \eta(x) = \sqrt{x} +\frac1{12} \frac{1}{\sqrt{x}} + O(x^{-3/2})
  \hspace{1cm}\mbox{ for }x\gg1
  \:.
\end{eqnarray}


\section{A perturbative approach to solve equation (\ref{thediffeq})
         \label{sec:pad}}

We solve equation (\ref{thediffeq}) with the boundary conditions 
(\ref{boundary}) in the limit $a\ll1$.
Let us write the solution as an expansion in powers of the parameter $a$~:
\begin{eqnarray}
  f(\chi) = f_0(\chi) + f_1(\chi) + f_2(\chi) + \cdots
  \:,
\end{eqnarray}
where $f_n(\chi)=O(a^n)$.
In order to satisfy the boundary conditions (\ref{boundary}) at any level
of approximation we impose $f_0(0)=1$ and $f_0(1)=0$ for the order 0, and 
$f_n(0)=f_n(1)=0$ for higher orders.
$f_0(\chi)$ is solution of $f_0''-bf_0=0$, therefore~:
\begin{eqnarray}
  f_0(\chi) = \frac{\sinh\sqrt{b}(1-\chi)}{\sinh\sqrt{b}}
  \:.
\end{eqnarray}
The first order term satisfies the differential equation~:
\begin{eqnarray}
  f_1''(\chi) - b f_1(\chi) = a \chi(1-\chi)f_0(\chi) \equiv \Psi(\chi)
  \:.
\end{eqnarray}
The solution satisfying the appropriate boundary conditions reads~:
\begin{eqnarray}
  f_1(\chi) &=& - \frac{1}{{\cal W}}
  \Bigg[ 
    f_0(\chi)
    \int_0^\chi\D t\:\Psi(t)\,f_0(1-t)
\nonumber\\
  &+& f_0(1-\chi)\int_\chi^1\D t\:\Psi(t)\,f_0(t)
  \Bigg]
  \:,
\end{eqnarray}
where ${\cal W}=\frac{\sqrt{b}}{\sinh\sqrt{b}}$ is the Wronskian of the 
two solutions ${\cal W}={\cal W}[f_0(\chi),f_0(1-\chi)]$.
After some calculations the derivatives are found~:
\begin{eqnarray}
  \label{fp0pepe}
  f'(0) &=& -\sqrt{b}\coth\sqrt{b} 
  - \frac{a}{4\sinh^2\sqrt{b}}
  \\\nonumber
  &&\times
  \left(
    -\frac13+\frac{\cosh^2\sqrt{b}}{b}-\frac{\sinh2\sqrt{b}}{2b^{3/2}}
  \right)+O(a^2)
  \\
  \label{fp1pepe}
  f'(1) &=& -\sqrt{b}\frac{1}{\sinh\sqrt{b}}
  + \frac{a}{4\sinh^2\sqrt{b}}
  \\\nonumber
  &&\times
  \left(
    \frac13\cosh\sqrt{b}-\frac{\cosh\sqrt{b}}{b}+\frac{\sinh\sqrt{b}}{b^{3/2}}
  \right)+O(a^2)
  \:.
\end{eqnarray}


\section{Relation between the weak localization and the conductivity 
         fluctuations\label{sec:genAB}}

We re-examine the relation between the weak localization and the 
conductivity fluctuations studied in ref.~\cite{AleBla02} for the case of 
the wire and used in ref.~\cite{LudMir04}. We show that the relation is 
more general and holds for the local conductivity
$\sigma=\int\frac{\D r\D r'}{{\rm Vol}}\sigma(r,r')$.
Note that it is only meaningful to consider a local conductivity
when the distribution of currents is uniform (translation invariant system
or a network with equal currents in its wires).

The weak localization is governed only by the phase coherence length
($L_N$ and/or $L_\varphi$). The study of conductivity fluctuations 
involves another important length scale~: the thermal length 
$L_T=\sqrt{D/T}$. Conductivity fluctuations are given by four
contributions~: the two first are interpreted as correlations 
of the diffusion constant~\cite{AltShk86}
\begin{eqnarray}
  \smean{\delta\sigma(B)\,\delta\sigma(B')}^{(1)}
  &\hspace{-0.1cm}= &\hspace{-0.1cm}
  \frac{16}{{\rm Vol}^2}\left(\frac{e^2}{h}\right)^2 
  \frac{\pi D^2}{3T}
  \int_0^\infty\D t\D t'\,\tilde\delta(t-t')
\nonumber\\ 
&&\hspace{-0.5cm}
  \times 
  \int\D r\D r'\,
  \widetilde{\cal P}_d(r,r';t)\,\widetilde{\cal P}_d(r,r';t')^*
\end{eqnarray}
where $\tilde\delta(t)=\frac{3T}{\pi}\left(\frac{\pi Tt}{\sinh\pi Tt}\right)^2$
is a function of width $1/T$ and normalized to unity.
The second contribution is obtained by replacing the diffuson by the cooperon 
$\widetilde{\cal P}_c$.
The two remaining contributions, of the form
$
\int_0^\infty\D t\D t'\,\tilde\delta(t-t')
\int\D r\D r'\,\re[\widetilde{\cal P}_d(r,r';t)\,\widetilde{\cal P}_d(r',r;t')]
$,
are interpreted as correlations of the density of states \cite{AltShk86}.
However, these two contributions are negligible \cite{AleBla02} since 
$L_T\ll L_N$ is necessary fulfilled ($L_T\sim L_N$ corresponds to the 
threshold of strong localization).
The condition $L_T\ll L_N$ allows us to replace the function $\tilde\delta(t)$
by $\delta(t)$ and obtain~:
\begin{eqnarray}
  \smean{\delta\sigma(B)\,\delta\sigma(B')}^{(1)}
  &=&\frac{16}{{\rm Vol}^2}\left(\frac{e^2}{h}\right)^2 \frac{\pi D^2}{3T}
  \nonumber\\
  && \hspace{-1.5cm} \times
  \int_0^\infty\D t\,
  \int\D r\D r'\,
  \left|\widetilde{\cal P}_d(r,r';t)\right|^2
  \:.
\end{eqnarray}
The diffuson and cooperon are solutions of the ``diffusion'' equation
\begin{widetext}
\begin{eqnarray}
  \left[
      \frac{\partial}{\partial t} 
   - D\left( \nabla-2\I e A_\mp \right)^2
   + \I\left(V_1(r,t)-V_2(r,t)\right)
  \right]
  \widetilde{\cal P}_{d,c}(r,r';t) = \delta(t)\,\delta(r-r')
  \:.
\end{eqnarray}
$A_\mp$ is the vector potential related to the magnetic field
$(B\mp B')/2$, where the sign is $-$ for $\widetilde{\cal P}_d$ and 
$+$ for $\widetilde{\cal P}_c$. The two potentials $V_1$ and $V_2$ are the two 
fluctuating electric potential associated to the two conductivity bubbles.
They are both characterized by the same fluctuations, given by the 
fluctuation-dissipation theorem (\ref{fdt}), however $V_1$ and $V_2$ are 
uncorrelated since they are associated to the conductivity bubbles
for two different configurations of the disorder~\cite{footnote6}~:
$\mean{V_i(r,t)\,V_j(r',t')}_V=\delta_{ij}
\frac{2e^2}{\sigma_0}T\,\delta(t-t')\,P_d(r,r')$.

Starting from the path integral representation of the diffuson it is possible
to perform the average over the fluctuating potential~: 
\begin{eqnarray}
  \mean{\left|\widetilde{\cal P}_d(r,r';t)\right|^2}_V 
  =
  \int_{r_1(0)=r'}^{r_1(t)=r}{\cal D}r_1(\tau)\,
  \int_{r_2(0)=r'}^{r_2(t)=r}{\cal D}r_2(\tau)\,
  \EXP{-\frac1{4D}\int_0^t\D\tau\,\dot r_1^2
     -\frac1{4D}\int_0^t\D\tau\,\dot r_2^2
     -\frac{4e^2T}{\sigma_0}\int_0^t\D\tau\, W(r_1(\tau),r_2(\tau))}
  \:,
\end{eqnarray}
where $W(r,r')$ was defined above by eq.~(\ref{defW}).
We introduce $r(\tau)$ defined for $\tau\in[0,2t]$ such that 
$r(\tau)=r_1(\tau)$ if $\tau\in[0,t]$ and $r(\tau)=r_2(2t-\tau)$ if 
$\tau\in[t,2t]$.
The the two path integrals can be gathered in one thanks to the integration 
over $r$~:
$
\int\D r\int_{r',0}^{r,t}{\cal D}r_1(\tau)
\int_{r',0}^{r,t}{\cal D}r_2(\tau)\rightarrow
\int_{r',0}^{r',2t}{\cal D}r(\tau)
$.
\begin{eqnarray}
  \int\D r
  \mean{\left|\widetilde{\cal P}_d(r,r';t)\right|^2}_V
  =\int_{r(0)=r'}^{r(2t)=r'}{\cal D}r(\tau)\,
  \EXP{-\frac1{4D}\int_0^{2t}\D\tau\,\dot r^2
       -\frac{2e^2T}{\sigma_0}\int_0^{2t}\D\tau\,W(r(\tau),r(2t-\tau))}
  \:.
\end{eqnarray}
\end{widetext}
This leads to the following relation~:
\begin{eqnarray}
  \label{AlBl1}
  \smean{\delta\sigma(B)\,\delta\sigma(B')}^{(1)}
  =-2\frac{e^2}{h}\frac{\pi}{3}\frac{L_T^2}{{\rm Vol}} 
  \smean{\Delta\sigma\left(\frac{B-B'}{2}\right)} 
\end{eqnarray}
where the weak localization is given by (\ref{spAAK2},\ref{spAAK2:cond}). 
Similarly the contribution of the cooperon gives~:
\begin{eqnarray}
  \label{AlBl2}
  \smean{\delta\sigma(B)\,\delta\sigma(B')}^{(2)}
  =-2\frac{e^2}{h}\frac{\pi}{3}\frac{L_T^2}{{\rm Vol}} 
  \smean{\Delta\sigma\left(\frac{B+B'}{2}\right)} 
  \:.
\end{eqnarray}
The total correlation function $\smean{\delta\sigma(B)\,\delta\sigma(B')}$ is
the sum of these two contributions. The relations (\ref{AlBl1},\ref{AlBl2})
were proved by Aleiner \& Blanter in ref.~\cite{AleBla02} by an explicit
calculation of the path integral for a wire and a plane, and comparison to the
result of AAK~\cite{AltAroKhm82}. Here we have demonstrated these relations
without having explicitly calculated the path integral, which makes our proof
more general, valid as soon as it is meaningful to consider a local
conductivity. The important physical consequence of these relations is that
the weak localization ({\it i.e.} the Altshuler--Aronov--Spivak oscillations)
and the conductance fluctuations ({\it i.e.} the Aharonov--Bohm oscillations)
are governed by the same length scale $L_N$. For example we expect the
amplitude of the AB oscillations to behave like
\begin{eqnarray}
  \delta g^{\rm AB}_{n} \propto  L_T\sqrt{L_N}
  \,\exp-|n|\frac{\pi}{16}\left(\frac{L}{L_N}\right)^{3/2}
\end{eqnarray}
for $L\gg L_N$.

\end{appendix}


\begin{thebibliography}{99}
\bibitem{footnote1}
  For the calculation of $\smean{\Delta\sigma}$,
  a factor 2 was missing in ref.~\cite{AltAroKhm82} and the correct factor can 
  be found in refs.~\cite{AleAltGer99,GouPiePotEstBir00,AkkMon04} for example.
\bibitem{footnote2}
  Note that the existence of a non exponential behaviour 
  $\EXP{-(t/\tau_N)^{3/2}}$ has been first noticed in ref.~\cite{SteAhaImr90}
  on qualitative ground.
\bibitem{footnote3}
  Note that LM have written their result in a different way~:
  $\langle \Delta \sigma_n \rangle \sim  \EXP{-|n|L/L_\phi^{AB}}$
  where  $L_\phi^{AB}$ depends on the 
  perimeter of the ring and is related to $L_N$ by 
  $L/L_\phi^{AB}=(L/L_N)^{3/2}\sim L^{3/2}T^{1/2}$. Therefore
  $L_\phi^{AB}\sim T^{-1/2}L^{-1/2}$.
  Throughout our article we  rather use the notation
  $L_c\equiv L_\phi^{AB}=L_N^{3/2}L^{-1/2}$.
\bibitem{footnote4} 
  It is interesting to note that AAK's result has been simultaneously derived
  by probabilists studying the statistical properties of the absolute area
  below a Brownian bridge (a Brownian motion conditioned to come back to its
  initial point) in ref.~\cite{She82} (moreover this author pointed out that
  the result was obtained even earlier in the context of economy in
  ref.~\cite{CifReg75}).  In this context eq.~(\ref{AAKAiry}) is interpreted
  as the double Laplace transform of the distribution below a Brownian bridge.
  The inverse Laplace transforms were performed in ref.~\cite{Ric82}.
\bibitem{footnote5}
  We stress that the temperature dependence of the harmonics $\EXP{-nT^{1/2}}$
  is characteristic of the diffusive regime (for large perimeter).  In the
  ballistic regime the harmonics of the AB oscillations behaves like
  $\EXP{-nT}$, as observed experimentally in ref.~\cite{HanKriPedSorLin01} and
  discussed theoretically in ref.~\cite{SeeBut01}.
\bibitem{footnote6}
  In a perturbative treatment of the electron-electron interaction 
  \cite{LeeRam85}, this 
  means that the e-e interaction lines do not couple the two conductivity
  bubbles.
\end{thebibliography}

\end{document}